\documentclass[aps, prc, twocolumn, floatfix, nofootinbib,
superscriptaddress, longbibliography, amsmath, amssymb]{revtex4-2}
\usepackage{amsfonts,amssymb,stmaryrd,latexsym,amsmath,braket}
\usepackage{appendix}
\usepackage{graphicx,subfigure}
\usepackage{comment}
\usepackage{times}
\usepackage{slashed}
\usepackage{bm}
\usepackage{braket}
\usepackage{hyperref}
\usepackage{cleveref}
\usepackage{mathrsfs}
\usepackage{qcircuit}
\usepackage{xcolor}

\graphicspath{{Figures/}}

\begin{document}

\title{Quantum state preparation by adiabatic evolution with customized gates}

\author{E. A. Coello P\'erez}
\email{coelloperez1@llnl.gov} 
\affiliation{Lawrence Livermore National Laboratory,
Livermore, CA 94550, USA}

\author{J. Bonitati}
\affiliation{Facility for Rare Isotope Beams, \& Department of Physics and Astronomy, Michigan State University 
East Lansing, MI 48824, USA}

\author{D. Lee}
\affiliation{Facility for Rare Isotope Beams, \& Department of Physics and Astronomy, Michigan State University 
East Lansing, MI 48824, USA}

\author{S. Quaglioni}
\affiliation{Lawrence Livermore National Laboratory,
Livermore, CA 94550, USA}

\author{K. A. Wendt}
\affiliation{Lawrence Livermore National Laboratory,
Livermore, CA 94550, USA}

\begin{abstract}

Quantum state preparation by adiabatic evolution is currently rendered ineffective by the long implementation times of the underlying quantum circuits, comparable to the decoherence time of present and near-term quantum devices.
These implementation times can be significantly reduced by realizing the evolution with a minimal number of customized gates. Employing a realistic model of a two-qubit processor, we carried out classical device-level simulations of the adiabatic evolution of a two-spin system implemented with customized two-qubit gates. These device-level simulations were compared with (experimental) ones solving the same problem on IBMQ systems.
When used to emulate the IBMQ quantum circuit, our device-level simulations reached state fidelities ranging from 65\% to 85\%, similar to the actual performance of a diverse set of IBMQ devices.
When we reduced the implementation time by using a minimal number of customized gates, however, the loss of fidelity was reduced by at least a factor of four, allowing us to accurately extract the energy of the target state. This improvement is enough to render adiabatic evolution useful for quantum state preparation for small systems or as a preconditioner for other state preparation methods.
\end{abstract}

\maketitle

\section{Introduction}
Quantum computing holds the key to efficiently solving problems that are intractable on classical computers~\cite{nielsen_ma2010,steane_a1998}. One such problem is that of simulating the dynamics of large quantum systems. While the resources required to solve this problem on classical computers grow exponentially with the size of the system, $N$, its solution on a universal quantum computer would require resources that grow only linearly in $N$, provided the evolution of the system is driven by local interactions~\cite{lloyd_s1996, feynman_rp1982}.
Before the simulation of an interesting system even starts, one must prepare the simulator on a state that corresponds to that of the system at the beginning of its evolution, often the ground state of the system's Hamiltonian. One method to prepare a universal quantum computer with arbitrary precision or fidelity is through adiabatic evolution from a state to which the device has easy access.
However, the universal quantum computer that evolves exclusively through controlled operations or quantum gates is just an ideal system. Preparation of present noisy intermediate-scale quantum (NISQ) computers by adiabatic evolution remains a challenge as the long implementation times required to achieve large fidelities enable interactions between the device and its environment to significantly deviate the evolution from its path to the target state.

In the digital quantum computing picture, quantum algorithms are typically implemented using a fixed basis of gates on sets of one and two qubits. Any many-qubit quantum algorithm can be implemented to high precision as a finite series of these gates~\cite{shende_vv2004, vatan_f2004, vidal_g2004, barenco_a1995}, however, the number of elementary gates required to implement complex quantum algorithms, such as adiabatic evolution, becomes prohibitively large. The short coherence times of modern devices therefore impose a harsh limit on the total run time of an algorithm before the resulting output is indistinguishable from noise. By extending that fixed basis of gates with a few custom gates that are ``aware'' of the details of the underlying quantum hardware and informed by the algorithm being implemented (e.g., details such as the time propagator between pairs of spins in a spin chain), highly optimized algorithms can be enacted. Often gates on physical quantum hardware are built from some analog signal applied to that quantum device (such as microwave pulses on superconducting transmons) and this approach can be thought of as performing a hybrid digital-analog quantum algorithm. Previously, this strategy has been applied successfully to simulate the real-time evolution of a system of neutron spins to sufficient duration and fidelity to extract spectroscopic information~\cite{wu2021, holland_et2020}.

Here, we present a noise-resilient approach for quantum state preparation using a minimal set of customized quantum gates to realize the adiabatic evolution of a two-spin system. We simulate the implementation of the adiabatic evolution using customized two-qubit gates by modeling a two-qubit processor as two capacitively coupled superconducting transmons driven by microwave pulses, and solving the Linblad master equation for its density matrix. We also carry out (experimental) digital simulations of the same adiabatic evolution on IBM Quantum (IBMQ) systems. We show that, when used to emulate a digital quantum simulation with execution times similar to those achieved on IBMQ, our two-qubit processor model yields state fidelities ranging from 65\% to 85\%, comparable with the experimental quantum simulations. However, by reducing the length of the microwave pulses realizing the customized gates and, consequently, the implementation time of the adiabatic evolution, we are able to reach the target state with up to 95\% fidelity. We further demonstrate that we can subsequently extract the energy of the target state, a significant improvement over the IBMQ simulations.

The paper is structured as follows. In Sec.~\ref{sec:adiabatic} we provide a brief review of adiabatic evolution, set up an algorithm evolving a system of two interacting spins, and discuss the two strategies we used to simulate the evolution of such system on quantum computers. The results of our simulations are described in Sec.~\ref{sec:implementation}. Finally, we discuss the results of this work and future research venues in Sec.~\ref{sec:conclusions}.

\section{Adiabatic evolution}
\label{sec:adiabatic}
A controllable quantum system can be slowly driven from the readily accessible ground state of an initial Hamiltonian $H_0$ into the ground state of an arbitrary Hamiltonian $H_T$ (encoding the dynamics of a desired system) by evolution with a time-dependent Hamiltonian
\begin{equation}
H(t) = f(t) H_0 + g(t) H_T,
\label{eq:Ht}
\end{equation}
where $f(t)$ and $g(t)$ are interpolation functions such that
\begin{equation}
\begin{gathered}
f(0) = 1 - g(0) = 1\\
{\rm and}\\
f(T) = 1 - g(T) = 0.
\end{gathered}
\end{equation}
This corresponds to imposing the boundary conditions $H(0)=H_0$ and $H(T)=H_T$. More in general, the adiabatic theorem states that a quantum system initially in the $k$-th eigenstate of $H_0$ will reach a state arbitrarily close to the $k$-th eigenstate of $H_T$ after a long enough evolution time $T$, provided the $k$-th eigenvalue is continuous throughout the evolution and does not cross other levels~\cite{born_m1928}. In terms of the parameter $s\equiv t/T$, the time $T$ required to prepare the quantum device with an error bounded from above by
\begin{equation}
\epsilon = || P(1) - P_k(1) ||,
\end{equation}
with $P(s)$ and $P_k(s)$ being, respectively, the projectors onto the evolved state and the $k$-th eigenstate of $H$ at time $s$, is of order~\cite{albash_t2018, boixo_s2010, boixo_s2009}
\begin{equation}
T \sim O\left(\max_s (|| \partial_s H(s) ||) / \Delta^2 \right),
\end{equation}
where $\Delta=\min_s(|\varepsilon_k(s)-\varepsilon_{k\pm1}(s)|)$ is the minimum energy gap involving the $k$-th eigenvalue throughout the evolution.

The quadratic dependence of the total evolution time $T$ on the inverse of the minimum energy gap, $1/\Delta$, represents a challenge for the use of adiabatic evolution as a method for quantum state preparation on current quantum devices, as well as those expected in the near future.
Long evolution times resulting from small energy gaps translate into implementation times long enough for these devices to lose coherence due to their interaction with the environment. 
Nevertheless, one strong motivation for improving the performance of adiabatic evolution is that it can serve as a preconditioner for other eigenstate preparation methods that require significant overlap between the initial state and the eigenstate of interest, such as phase estimation~\cite{Kitaev:1995qy} or the rodeo algorithm~\cite{Choi:2020pdg, Qian:2021wya}. Even noisy or incomplete adiabatic evolution can provide a large enhancement of the initial state overlap, thereby significantly improving the performance of the quantum state preparation algorithm applied thereafter. For example, improving the initial state overlap probability from $0.1\%$ to $5\%$ would provide a fifty-fold improvement in the algorithmic efficiency.

\subsection{Adiabatic evolution of two-spin systems}
\label{sec:adb2spin}
\begin{figure}[t]
\centering
\includegraphics[width=\columnwidth]{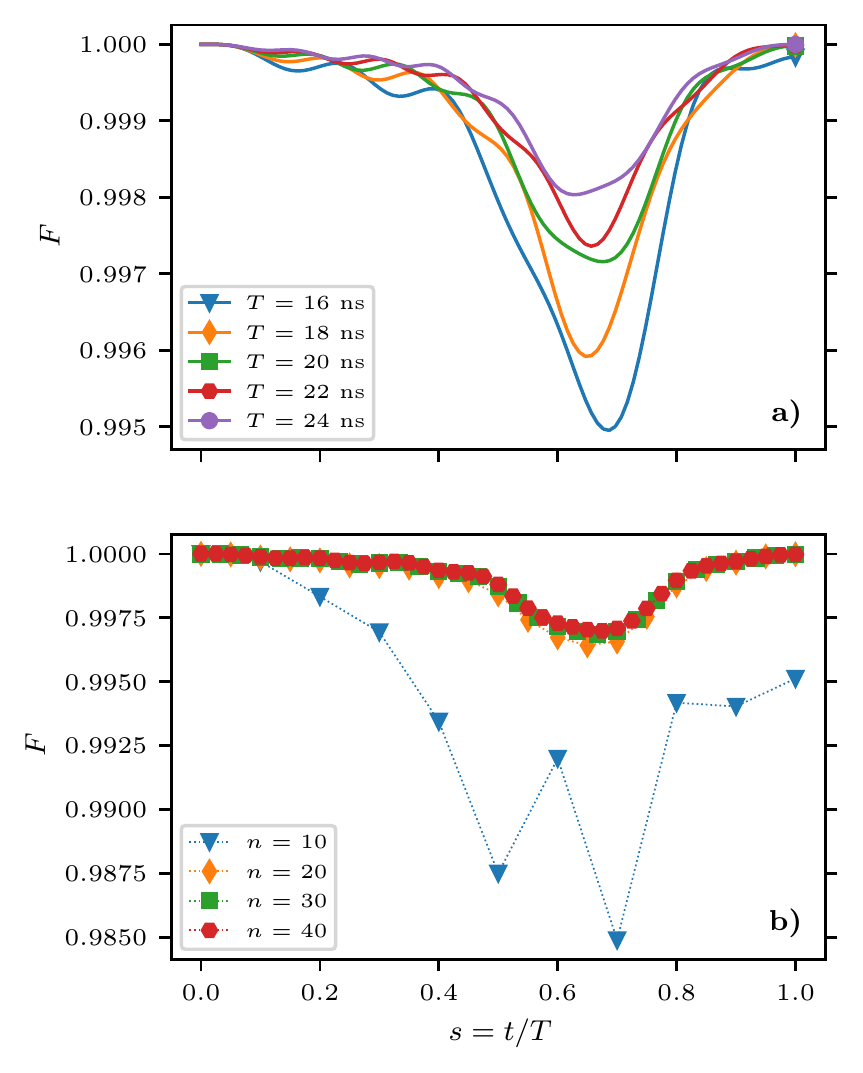}
\caption{Evolution of the fidelity between the state of the device and the ground state of $H(t)$ for different values of {\bf{a)}} the evolution time $T$ and {\bf{b)}} the number of time steps $n$ as a function of the parameter $s\equiv t/T$. The adiabatic evolution we implement on quantum devices uses $T=20$~ns and $n=20$, as this combination yields negligible infidelity $1-F$ at the end of the evolution while minimizing the number of gates required for its implementation.}
\label{fig:exactevolution}
\end{figure}

In the present study we consider the preparation of a two-spin system in the ground state of the Hamiltonian
\begin{align}
H_T = - \sigma^x_1\sigma^x_2 + \sigma^y_1\sigma^y_2 +
\frac{1}{2} \sigma^z_1 \sigma^z_2 - \sum_{i=1}^2 \sigma^z_i\,,
\label{eq:HT}
\end{align}
by initializing the system in the ground state of
\begin{align}
H_0 = \sum_{i=1}^2 \sigma^x_i,
\label{eq:H0}
\end{align}
and performing the adiabatic evolution imposed by the time-dependent Hamiltonian of Eq.~\eqref{eq:Ht} with interpolation functions
\begin{align}
f(t)=\cos^2(\pi t/2T) & & g(t)=1-f(t).
\label{eq:interpolation}
\end{align}
We note that the ground state of $H_0$ is a linear combination of the uncoupled two-spin states,
\begin{equation}
    \ket{\phi(0)}
    =\frac{1}{2}\left(\ket{\downarrow\downarrow}-\ket{\downarrow\uparrow}-\ket{\uparrow\downarrow}+\ket{\uparrow\uparrow}\right).
\end{equation}
The ground state of $H_T$,
\begin{equation}
    \ket{\phi(T)}
    =\mathscr{N}\left[\left(-1+\sqrt{2}\right)\ket{\downarrow\downarrow}+\ket{\uparrow\uparrow}\right],
\end{equation}
with $\mathscr{N}$ a normalization constant, has energy $E_T=-2.328$. Similarly, we can introduce the instantaneous ground state $\ket{\phi(t)}$ of $H(t)$ by solving the time-independent Schr\"odinger equation $H(t)\ket{\phi(t)}=E_t\ket{\phi(t)}$ at each instant $t$.  
The solution of the time-dependent Schr\"odinger equation $i\frac{d}{dt}\ket{\psi(t)}=H(t)\ket{\psi(t)}$ will then approximate $\ket{\phi(t)}$ with (instantaneous) fidelity
\begin{equation}
F(t) = |\langle \phi(t) | \psi(t) \rangle|.
\label{eq:fid}
\end{equation}
For a sufficiently long evolution time $T$, the state of the system at time $T$ will be very close to the ground state of $H_T$ [see Fig.~\ref{fig:exactevolution}(a)].
Even for a relatively short time of $T=16$~ns, the infidelity $1-F(T)$ at the end of the evolution is bounded from above by $10^{-3}$. In what follows we set $T=20$~ns. This choice is a good compromise in anticipation of minimizing, as much as possible, the run time (and hence decoherence error) that one would face when implementing this adiabatic evolution on present-day quantum devices. 

The adiabatically evolved state can be expressed in terms of the unitary-time evolution operator for a time-dependent Hamiltonian,
\begin{align}
\label{eq:U}
    \ket{\psi(T)} &= \mathcal{U}(0,T)\ket{\psi(0)}\\\nonumber
    &=\mathcal{T}\exp\left(-\frac{i}{\hbar} \int_0^T H(\tau) d\tau \right) \ket{\psi(0)}\,,
\end{align}
where $\mathcal{T}\exp$ denotes the time-ordered exponential. To implement adiabatic evolution on a quantum device it is useful to divide the evolution time into $n$ steps and approximate the evolution operator $\mathcal{U}(0,T)$ as the product of $n$ short-time propagators (for the $n$ instantaneous Hamiltonians)
\begin{equation}
\mathcal{U}(0,T) \approx \prod_{k=1}^{n} U(t_k) =
\prod_{k=1}^{n} e^{-iH(t_k)\Delta t},
\label{eq:stp}
\end{equation}
with $\Delta t=T/n$. The error introduced by the discretization of the evolution time increases with the variation with time of the Hamiltonian $dH(t)/dt$, and is well behaved provided $H(t_k)$ is a good approximation to the average of $H(t)$ in the time interval $[t_{k-1},t_k]$~\cite{poulin_d2011}.
Thus, the discretization error can be controlled by the number of steps $n$ in which the evolution time is divided [see Fig.~\ref{fig:exactevolution}(b), where the fidelity is now evaluated with respect to the solutions $\ket{\psi(t)}$ obtained from the unitary-time evolution of Eq.~\eqref{eq:U} in the approximation of Eq.~\eqref{eq:stp}].
While larger $n$ values guarantee smaller discretization errors, fewer steps are preferable to minimize the number of quantum gates (circuit depth) required to implement the algorithm on noisy intermediate-scale quantum (NISQ) devices and hence reduce loss of coherence due to interactions with the environment.
The results shown in Fig.~\ref{fig:exactevolution}(b) indicate that the discretization error becomes negligible with as few as 20 time steps. Therefore, in what follows we set $n=20$.

\subsection{Implementation on two-qubit devices}
\label{sec:implementation}
Performing a quantum simulation of the adiabatic evolution of Sec.\ref{sec:adb2spin} requires mapping the spin degrees of freedom to the states of a quantum processors, and translating the short-time propagators $U(t_k)$ in Eq.~\eqref{eq:stp} into quantum gates. 
The uncoupled states of the two-spin system can be trivially mapped to the computational states of a two-qubit system. Specifically, we use the $|00\rangle$, $|01\rangle$, $|10\rangle$ and $|11\rangle$ states to represent, respectively, the uncoupled two-spin states $\ket{\downarrow\downarrow}$, $\ket{\downarrow\uparrow}$, $\ket{\uparrow\downarrow}$ and $\ket{\uparrow\uparrow}$.
Concerning the translation of the short-time propagators into quantum gates, we consider two strategies: $(i)$ a standard decomposition of each propagator into a circuit of elementary gates; and $(ii)$ a direct implementation as a single customized two-qubit gate.

\begin{figure}[b]
\centering
\begin{flushleft}{\quad \normalsize{\bf{a)}}}\end{flushleft}\[
\begin{array}{c}
\Qcircuit @C=1.5em @R=1.5em {
& \multigate{1}{U(t_1)} & \qw & \cdots & & \multigate{1}{U(t_k)} & \qw & \cdots & & \multigate{1}{U(t_n)} & \qw \\
& \ghost{U(t_0)} & \qw & \cdots & & \ghost{U(t_k)} & \qw & \cdots & & \ghost{U(t_k)} & \qw \gategroup{1}{6}{2}{6}{1.0em}{.}
}
\end{array}\]
\begin{flushleft}
\quad \normalsize{\bf{b)}} \qquad\qquad\qquad\qquad\qquad
$\Big\Updownarrow$
\end{flushleft}\[
\begin{array}{c}
\Qcircuit @C=0.75em @R=1.5em {
\cdots & & \gate{{\rm U3}({\boldsymbol{\alpha}}_{k})} & \ctrl{1} & \gate{{\rm U3}({\boldsymbol{\beta}}_{k})} & \ctrl{1} & \gate{{\rm U3}({\boldsymbol{\gamma}}_{k})} & \ctrl{1} & \gate{{\rm U3}({\boldsymbol{\delta}}_{k})} & \qw & \cdots \\
\cdots & & \gate{{\rm U3}({\boldsymbol{\epsilon}}_{k})} & \targ & \gate{{\rm U3}({\boldsymbol{\zeta}}_{k})} & \targ & \gate{{\rm U3}({\boldsymbol{\eta}}_{k})} & \targ & \gate{{\rm U3}({\boldsymbol{\theta}}_{k})} & \qw & \cdots \gategroup{1}{3}{2}{9}{1.0em}{.}
}
\end{array}\]
\caption{Decomposition of the $k$-th short-time propagator in the adiabatic evolution algorithm [circuit {\bf{a)}}] into CNOT and U3 gates [circuit {\bf{b)}}]. Each U3 gate depends on three Euler angles. The circuit resulting from this decomposition can be directly implemented on IBMQ systems.}
\label{fig:ibmqdecomp}
\end{figure}
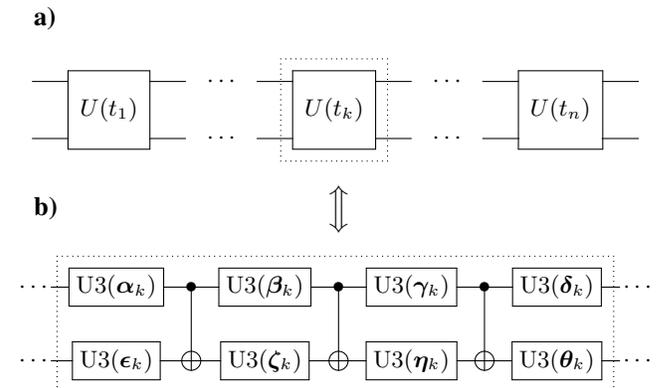

The first strategy exploits the fact that any unitary operation involving two qubits, such as the short-time propagators $U(t_k)$ in Eq.~\eqref{eq:stp}, can be expressed in terms of three CNOT gates and eight one-qubit U3 gates (constructed from fifteen $x$, $y$ or $z$ one-qubit rotations)~\cite{shende_vv2004,vatan_f2004,vidal_g2004} as schematically shown in Fig.~\ref{fig:ibmqdecomp}.
Such a decomposition can be easily implemented on cloud quantum computing platforms. We studied the performance of this approach by running simulations on several IBMQ systems~\cite{ibmq_2021}.
The quantum circuit implementing the adiabatic-evolution algorithm was build using the open-source Quantum information software kit (Qiskit)~\cite{qiskit_2021}, which allowed us to decompose each short-time propagator into elementary gates with the build-in function \textit{quantum\_info.two\_qubit\_cnot\_decompose}. The decomposition of the first short-time propagator is explicitly shown in Appendix~\ref{sec:decomposition} as an example.
The results of quantum simulations on IBMQ processors employing this algorithm are discussed in Sec.~\ref{sec:strategy1}.

\begin{figure}
\centering
\includegraphics[width=\columnwidth]{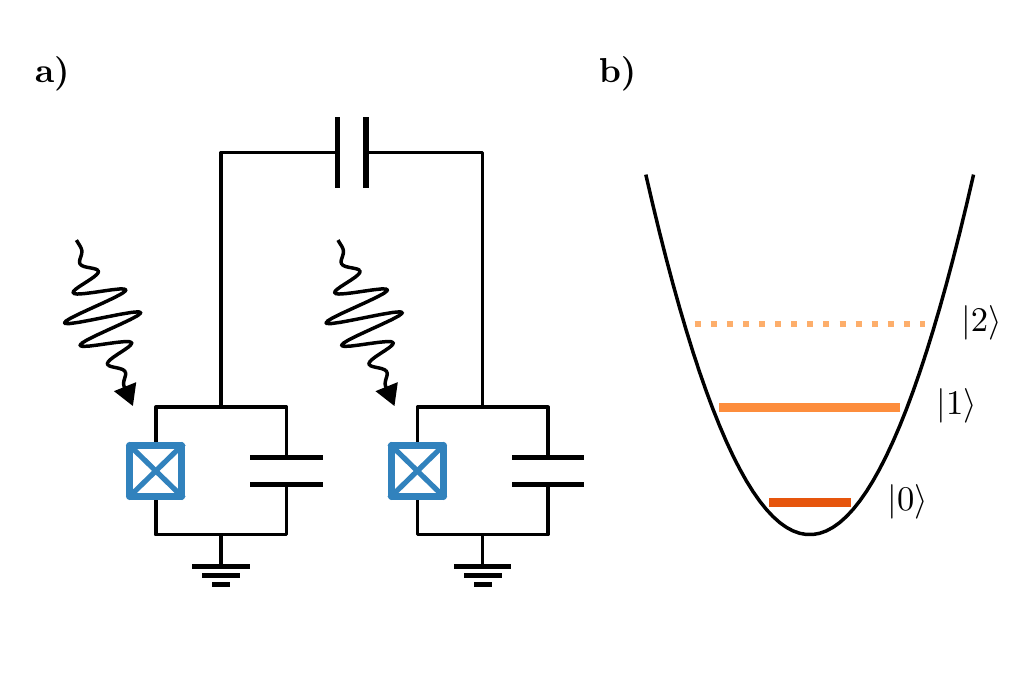}
\caption{Schematic representation of {\bf{a)}} the model two-qubit processor, consisting of two capacitively coupled superconducting transmons controlled with microwave pulses, and {\bf{b)}} the energy spectrum of a superconducting transmon in which the lowest two levels, shown as solid lines, are the computational qubit.}
\label{fig:modelprocessor}
\end{figure}

The second strategy employs a realistic model of a physical quantum device to realize each short-time propagator in Eq.~\eqref{eq:stp} with a single customized gate. In the present study, we model a two-qubit processor as two capacitively coupled superconducting transmons controlled by microwave pulses, as schematically shown in Fig.~\ref{fig:modelprocessor} {\bf{a)}}.
The dynamics of this system is described by the Hamiltonian (see Appendix~\ref{sec:qpumodel} for details)
\begin{equation}
H_{\rm QPU}(t) = H_{\rm d} + H_{\rm c}(t),
\label{eq:HQPU}
\end{equation}
where
\begin{equation}
H_{\rm d} \approx
-\sum_{i=1}^2 \alpha a_i^{\dagger}a_i a_i^{\dagger}a_i -
g \left( a_1^\dagger a_2 + a_2^\dagger a_1 \right)
\end{equation}
is the drift Hamiltonian of the unperturbed device, and 
\begin{equation}
H_{\rm c}(t) = \sum_{i=1}^2 \left[
\epsilon_{\rm I}^i(t) (a_i^\dagger + a_i) -
i \epsilon_{\rm Q}^i(t) (a_i^\dagger - a_i) \right]
\end{equation}
is the time-dependent Hamiltonian describing the control of the quantum processor by irradiation with resonant microwave pulses.
In the expressions above, $a^\dagger_i$ and $a_i$ are the creation and annihilation operators of transmon $i$, $\alpha=200$~MHz is the anharmonicity of both transmons, and $g=3$~MHz is the strength of the interaction or crosstalk between the transmons due to the capacitive coupling.
The time-dependent amplitudes $\epsilon_{\rm I}^i(t)$ and $\epsilon_{\rm Q}^i(t)$ are, respectively, the in-phase and quadrature tunable pulse sequences controlling transmon $i$. In the present study, we take into account the first three energy levels of each transmon. The subspace of states with zero and one quanta per transmon define the computational two-qubit states. The explicit inclusion of states with two quanta in at least one of the transmons provides a description of higher-energy states that, in a realistic quantum simulation, can be populated due to gate error and decoherence. It also allows our control pulses to reduce this leakage by blocking transitions to these states, as similarly done by the derivative removal by adiabatic gate (DRAG) algorithm~\cite{motzoi_f2009}.

\begin{figure}
\centering
\includegraphics[width=\columnwidth]{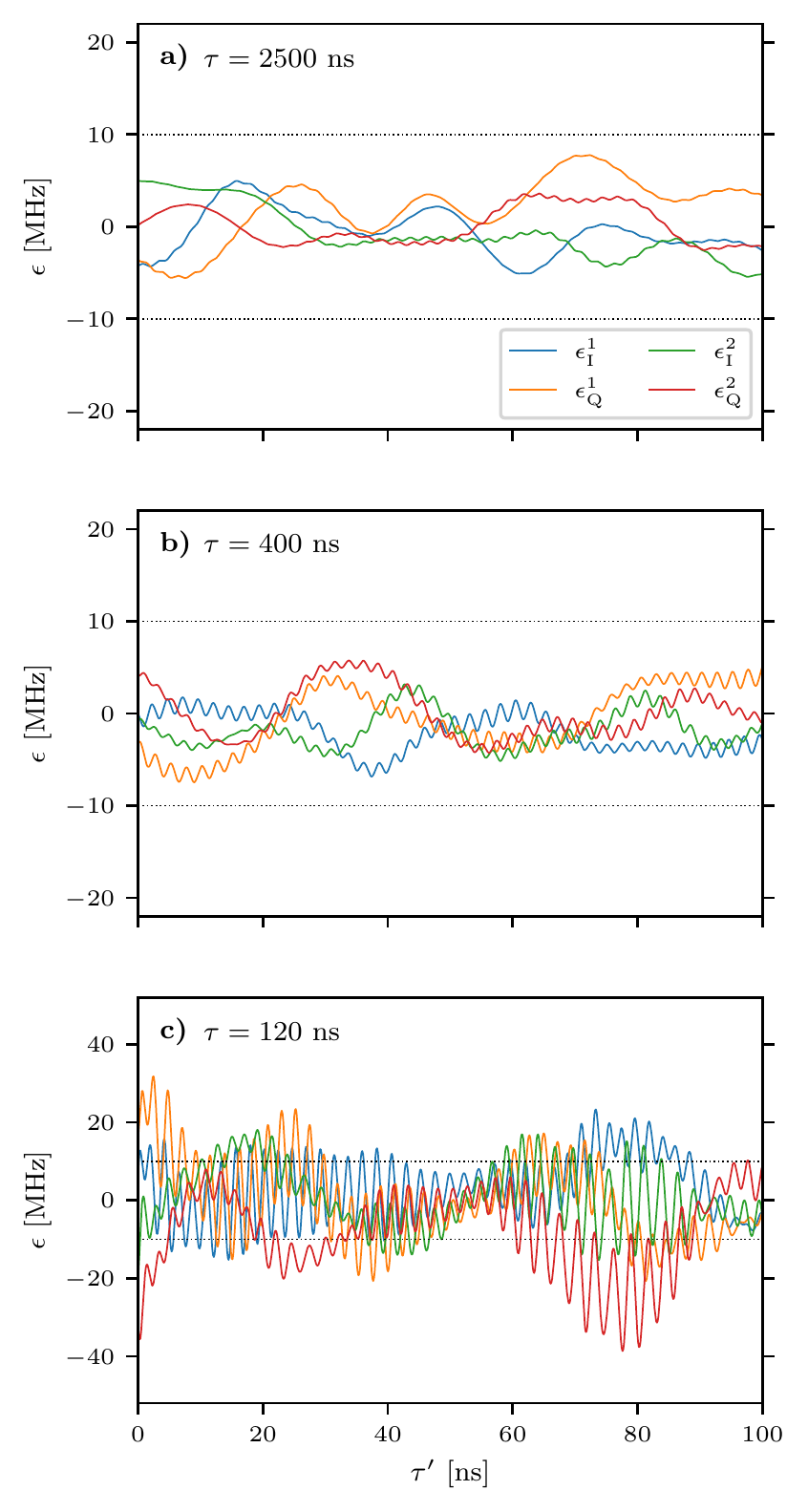}
\caption{First 100~ns of the control pulses with lengths {\bf{a)}} $\tau=2500$~ns, {\bf{b)}} $\tau=400$~ns and {\bf{c)}} $\tau=120$~ns realizing the first short-time propagator in the adiabatic-evolution algorithm. The dependence of the amplitude of the control pulse on its length sets a lower bound for the implementation time of two-qubit gates on present-day quantum devices.}
\label{fig:controlpulses}
\end{figure}

We compute the customized two-qubit gates realizing each of the short-time propagators in Eq.~\eqref{eq:stp} by finding the pulse sequences $\epsilon_{\rm I}^i(t)$ and $\epsilon_{\rm Q}^i(t)$ that solve (within an acceptable accuracy) the optimization problem
\begin{align}
\label{eq:QOC}
U_{\rm QPU}(t_k) &\simeq {\mathcal U}_{\rm QPU}(0,\tau)\\
& = \mathcal{T} \exp\left( -\frac{i}{\hbar} \int_0^\tau
\left[H_{\rm d} + H_{\rm c}(\tau') \right] d\tau' \right),\nonumber
\end{align}
where $U_{\rm QPU}(t_k)$ represents the short-time propagator $U(t_k)$ embedded into the Hilbert space spanned by the considered two-transmon states. Employing the gradient ascent pulse engineering (GRAPE) algorithm~\cite{khaneja_n2005}, the solution to Eq.~\eqref{eq:QOC} is found by minimizing the objective function
\begin{equation}
\Phi = 1 - \frac{F_{\rm gate}^2}{2} + \chi \frac{\exp(\bar{\epsilon}^{2n})-1}{\exp(1)-1}\,,
\label{eq:of}
\end{equation}
where
\begin{equation}
F_{\rm gate} = \left|\frac{{\rm tr}(U_{\rm QPU}^\dagger\mathcal{U}_{\rm QPU})}
{{\rm dim}_{\rm QPU}}\right|,
\end{equation}
with ${\rm dim}_{\rm QPU}$ the dimension of the considered Hilbert space, and
\begin{equation}
\bar{\epsilon} = \frac{1}{\epsilon_{\rm cut}}
\sqrt{
\frac{1}{T} \sum_{i=1}^2\sum_{j\in\{{\rm I, Q}\}} \int_0^T \epsilon^i_j(t)^2 dt }
\end{equation}
the root-mean-squared amplitude of the control pulse normalized to $\epsilon_{\rm cut}$. The gate fidelity $F_{\rm gate}$ in the second term on the right-hand-side of Eq.~\eqref{eq:of} provides an indicator of the accuracy with which the desired unitary operation is reproduced.  
The last term penalizes large amplitudes through the parameters $\epsilon_{\rm cut}$, which describes the cutoff amplitude, and $n$, which sets the harshness of the cutoff, all with relevance dictated by the parameter $\chi$. The introduction of this penalty allows one to avoid high-amplitude solutions were the approximation for the Hamiltonian entering the optimization of Eq.~\eqref{eq:QOC} is no longer valid, and where hardware realizing the control pulses might not work optimally. 
The form of this loss term is chosen such that a zero amplitude pulse yields zero contribution to the total loss and that an pulse whose root-mean-square amplitude is $\epsilon_{\textrm{cut}}$ contributes $\chi$ to the total loss.
Fig.~\ref{fig:controlpulses} shows the first 100~ns of control pulses realizing the first short-time propagator in the adiabatic evolution algorithm obtained for three different pulse lengths $\tau$ of 2500~ns, 400~ns and 120~ns [see Eq.~\eqref{eq:QOC}] and a sampling rate of 8 samples per ns.
The pulse length $\tau=2500$~ns was chosen to approximately match the implementation time $\tau_U$ of the short-time propagators on the IBMQ Belem system (see Table~\ref{table:IBMQspecs}), which will allow us to compare our device-level simulations to quantum simulations on that system. The pulse length $\tau=400$ is representative of CNOT implementation times on IBMQ systems, while pulses of length $\tau=120$~ns where the shortest for which the gate infidelity, $1-F_{\rm gate}$, could be kept below $10^{-4}$.
All control pulses were found by minimizing the objective function of Eq.~\eqref{eq:of} with $\epsilon_{\rm cut}=30$~MHz, $n=3$, and $\chi=10^{-3}$.
Decreasing the length of the control pulse increases the root-mean-squared amplitude as well as the relevance of its high-frequency components. This dependence of the amplitude of the control pulse on its length sets a lower bound for the implementation time of two-qubit gates, and hence on the overall implementation time for the whole evolution on present-day quantum devices, as $(i)$ approximations for the Hamiltonian of a quantum computer [used to solve Eq.~\eqref{eq:QOC}] and $(ii)$ hardware employed to control it work optimally only within some energy regime.

To study the performance of the adiabatic evolution implemented by means of customized gates obtained through the minimization of Eq.~\eqref{eq:of}, we carried out classical device-level simulations of the evolution of the model two-transmon processor. The results of these classical simulations are discussed in Sec.~\ref{sec:strategy2}

\section{Results}

\subsection{Adiabatic evolution of two-spin system on IBMQ}
\label{sec:strategy1}
We ran experiments on IBMQ systems to assess how viable it is to simulate the adiabatic evolution of Eq.~\eqref{eq:stp} on them using circuits of elementary gates. The calibration data for the used IBMQ systems are listed in Table~\ref{table:IBMQspecs}.
\begin{table}[b]
\centering
\caption{Calibration data for \textit{ibmq\_belem}, \textit{casablanca}, \textit{ibmq\_lima} and \textit{ibmq\_manila}. The listed relaxation and dephasing times, $T_1$ and $T_2$, are those reported for qubits 0 and 1 of the corresponding system throughout the simulation. Implementation times for the CNOT gate and the decomposition of the short-time propagators are respectively listed as $\tau_{\rm CNOT}$ and $\tau_U$.}
\begin{tabular}{c|c|c|c|c|c|c}
\hline\hline
IBMQ system & \multicolumn{2}{c|}{$T_1$ [$\mu$s]} & \multicolumn{2}{c|}{$T_2$ [$\mu$s]} &
$\tau_{\rm CNOT}$ [ns] & $\tau_{U}$ [ns] \\
\hline
\textit{ibmq\_belem} & 102.6 & 70.4 & 127.3 & 104.5 & 810.7 & $\approx$2500 \\
\textit{ibmq\_casablanca} & 111.7 & 130.1 & 40.7 & 102.2 & 760.9 & $\approx$2400 \\
\textit{ibmq\_lima} & 101.6 & 113.0 & 180.0 & 106.9 & 305.8 & $\approx$1000 \\
\textit{ibmq\_manila} & 136.0 & 244.2 & 112.8 & 46.7 & 277.3 & $\approx$900 \\
\hline\hline
\end{tabular}
\label{table:IBMQspecs}
\end{table}
We started the simulations by initializing the IBMQ processors in the ground state of the initial Hamiltonian $H_0$ [see Eq.~\eqref{eq:H0}] by applying the two-qubit Pauli $X^{(2)}= \sigma^x_1\sigma^x_2$ and Hadamard $H^{(2)}= H_1H_2$ gates to the qubits' lowest state,
\begin{equation}
\ket{\psi(0)}=H^{(2)}X^{(2)}\ket{00}
=\frac{1}{2}\left(\ket{00}-\ket{01}-\ket{10}+\ket{11}\right),
\label{eq:psi0}
\end{equation}
and carried out the adiabatic evolution by applying the circuit of elementary quantum gates resulting from the implementation of the $n$ short-time propagators discussed in Sec.~\ref{sec:implementation}.

For each considered IBMQ system, we ran several simulations allowing us to approximately reconstruct the evolution of the instantaneous fidelity [see Eq.~\eqref{eq:fid}] and the expectation value of the target Hamiltonian $H_T$ [see Eq.~\eqref{eq:HT}],
\begin{equation}
\langle H_T \rangle(t)=\bra{\psi(t)}H_T\ket{\psi(t)},
\label{eq:<HT>}
\end{equation}
using quantum expectation estimation (for details see Appendix~\ref{sec:tomography}).
In Fig.~\ref{fig:elementaryevo} we compare the results from simulations on IBMQ systems with those from simulations on an ideal quantum processor, carried out with Qiskit's Aer simulator.
\begin{figure}[t]
\centering
\includegraphics[width=\columnwidth]{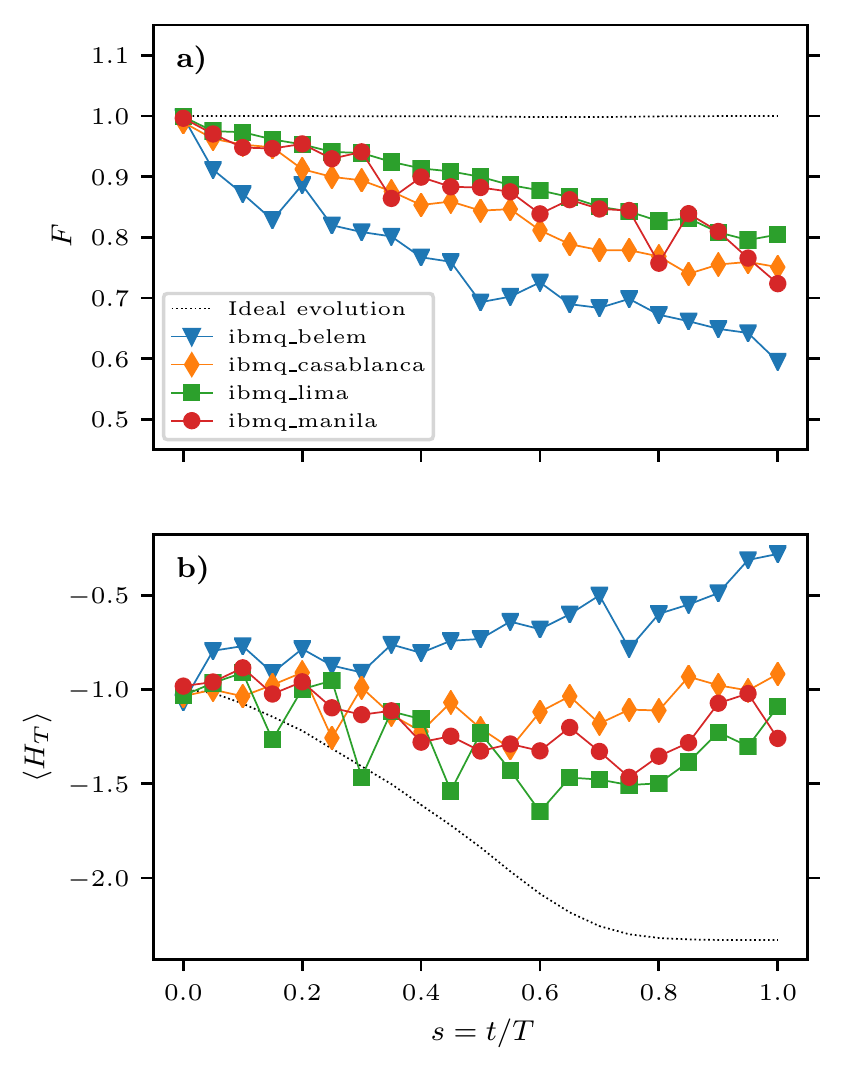}
\caption{Evolution of {\bf{a)}} the instantaneous fidelity $F(t)$ and {\bf{b)}} the expectation value $\braket{H_T}(t)$ resulting from simulations of the adiabatic evolution of Eq.~\eqref{eq:stp} on \textit{ibmq\_belem} (blue triangles), \textit{ibmq\_casablanca} (orange diamonds), \textit{ibmq\_lima} (green squares) and \textit{ibmq\_manila} (red circles). Large deviations from simulations on an ideal quantum processor carried out with Qiskit's Aer simulator (dotted lines) result from loss of coherence due to long implementation times.}
\label{fig:elementaryevo}
\end{figure}
The ideal results, shown as dotted lines, are nearly indistinguishable from the exact results (shown in Fig~\ref{fig:exactevolution}). This demonstrates that, for two-qubit systems, gate error does not degrade significantly the fidelity with which the target state is reached.
On the other hand, quantum simulations on \textit{ibmq\_belem}, \textit{ibmq\_casablanca}, \textit{ibmq\_lima} and \textit{ibmq\_manila}, shown as color markers, deviate from the ideal evolution reaching the target state with fidelities of 60\%, 75\%, 80\% and 72\%, respectively. The loss of fidelity throughout the evolution caused the properties extracted from the reached state to be significantly different from those of the ground state of the target Hamiltonian, as shown by the evolution of $\braket{H_T}$ in Fig.~\ref{fig:elementaryevo} {\bf{b)}}. The energies extracted from the IBMQ simulations range in mean value from $-1.25$ to $-0.25$, whereas the ideal simulation reaches the target value $E_T=-2.328$.
We notice that the rate at which fidelity is lost remains approximately constant throughout the evolution. This is somehow surprising, as our choice for the interpolation functions defining $H(t)$ [see Eq.~\eqref{eq:interpolation}] is such that the evolution is ``slowest'' at its beginning and end points (see Fig.~\ref{fig:exactevolution}), where we expected small fidelity losses. The constant rate at which fidelity is lost suggests this effect is dominated by either the gate error accumulated from the elementary gates realizing the short time propagator (for which the largest contributions come from errors in the CNOT gates) or the loss of coherence along their combined implementation time.

These results evince the main challenge of quantum state preparation of NISQ devices by means of adiabatic evolution: The implementation time of the quantum circuit expressing the adiabatic evolution of Eq.~\eqref{eq:stp} in terms of elementary gates is comparable to the decoherence times of present quantum devices.
Specifically for this work, the implementation times of the adiabatic evolution on \textit{ibmq\_belem}, \textit{ibmq\_casablanca}, \textit{ibmq\_lima} and \textit{ibmq\_manila}, respectively approximated as 52.5~$\mu$s, 50.4~$\mu$s, 21~$\mu$s and 18.9~$\mu$s, are comparable to the relaxation and dephasing decoherence times of these systems, listed as $T_1$ and $T_2$ in Table~\ref{table:IBMQspecs}.

Besides errors due to loss of coherence, simulations on NISQ devices carry measurement errors. These are mitigated in this work by inverting the confusion matrix, extracted from calibration experiments performed at each time step. (For details on the measurement errors and their mitigation see Appendix~\ref{sec:mitigation}.)

\subsection{Classical device-level simulations with customized gates}
\label{sec:strategy2}
An alternative to implementing the adiabatic evolution through elementary gates is to find pulse sequences realizing the short-time propagators (see Sec.\ref{sec:implementation}), aiming to reduce either the evolution's circuit depth or implementation time or both.
To study the performance of this alternative approach, we carried out classical device-level simulations of the evolution of the model two-transmon processor (see Sec.~\ref{sec:implementation}). Specifically, we used the Quantum Toolbox in Python (QuTiP)~\cite{johansson_jr2013, johansson_jr2012} to solve the Linblad master equation
\begin{equation}
\begin{aligned}
\dot{\rho} =& -\frac{i}{\hbar}
\left[ H_{\rm QPU}, \rho \right]\\
&+ \frac{1}{T_1} \sum_{i=1}^2 \left( a_i\rho a^\dagger_i
- \frac{1}{2}\left\{a_i a^\dagger_i,\rho\right\} \right)\\
&+ \frac{1}{T_2} \sum_{i=1}^2
\left( a^\dagger_i a_i\rho a_i a^\dagger_i - \frac{1}{2}
\left\{a_i a^\dagger_i a^\dagger_i a_i,\rho\right\} \right)\,,
\end{aligned}
\end{equation}
for the density $\rho$ of the two-transom system. Here, $H_{\rm QPU}$ is the two-transmon Hamiltonian introduced in Eq.~\eqref{eq:HQPU} with control pulses optimized to realize the short-time propagators in Eq.~\eqref{eq:stp}.
The decoherence mechanisms considered by the master equation, relaxation and dephasing, are parameterized by the decoherence times $T_1$ and $T_2$, respectively. Our classical device-level simulations take values for these parameters from calibration data of the IBMQ devices on which we conducted experiments [see Table~\ref{table:IBMQspecs}], to emulate the interaction of these systems with the environment.

\begin{figure}[t]
\centering
\begin{flushleft}{\quad \normalsize{\bf{a)}}}\end{flushleft}\[
\begin{array}{c}
\Qcircuit @C=0.75em @R=1.5em {
\cdots & & \gate{{\rm U3}({\boldsymbol{\alpha}}_{k})} & \ctrl{1} & \gate{{\rm U3}({\boldsymbol{\beta}}_{k})} & \ctrl{1} & \gate{{\rm U3}({\boldsymbol{\gamma}}_{k})} & \ctrl{1} & \gate{{\rm U3}({\boldsymbol{\delta}}_{k})} & \qw & \cdots \\
\cdots & & \gate{{\rm U3}({\boldsymbol{\epsilon}}_{k})} & \targ & \gate{{\rm U3}({\boldsymbol{\zeta}}_{k})} & \targ & \gate{{\rm U3}({\boldsymbol{\eta}}_{k})} & \targ & \gate{{\rm U3}({\boldsymbol{\theta}}_{k})} & \qw & \cdots \gategroup{1}{5}{2}{5}{1.0em}{.}
}
\end{array}\]
\begin{flushleft}
\quad \normalsize{\bf{b)}} \qquad\qquad\qquad\qquad\qquad
$\Big\Updownarrow$
\end{flushleft}\[
\begin{array}{c}
\Qcircuit @C=0.75em @R=1.5em {
\cdots & & \multigate{1}{{\rm U3}^{(2)}} & \ctrl{1} & \multigate{1}{{\rm U3}^{(2)}} & \ctrl{1} & \multigate{1}{{\rm U3}^{(2)}} & \ctrl{1} & \multigate{1}{{\rm U3}^{(2)}} & \qw & \cdots \\
\cdots & & \ghost{{\rm U3}^{(2)}} & \targ & \ghost{{\rm U3}^{(2)}} & \targ & \ghost{{\rm U3}^{(2)}} & \targ & \ghost{{\rm U3}^{(2)}} & \qw & \cdots \gategroup{1}{5}{2}{5}{1.0em}{.}
}
\end{array}\]
\caption{Combination of simultaneous two U3 gates [circuit {\bf{a)}}] into two-qubit ${\rm U}^{(2)}$ gates [circuit {\bf{b)}}]. The CNOT and ${\rm U3}^{(2)}$ gates in circuit {\bf{b)}} are realized with control pulses which combined lengths match the implementation time of circuit {\bf{a)}} on IBMQ systems.}
\label{fig:directcomparison}
\end{figure}
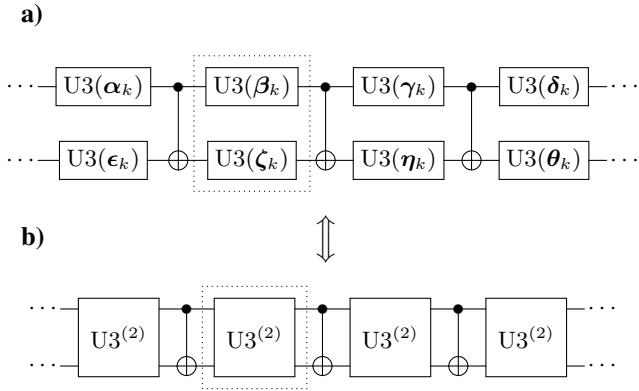

\begin{figure*}[ht]
\centering
\includegraphics[width=\textwidth]{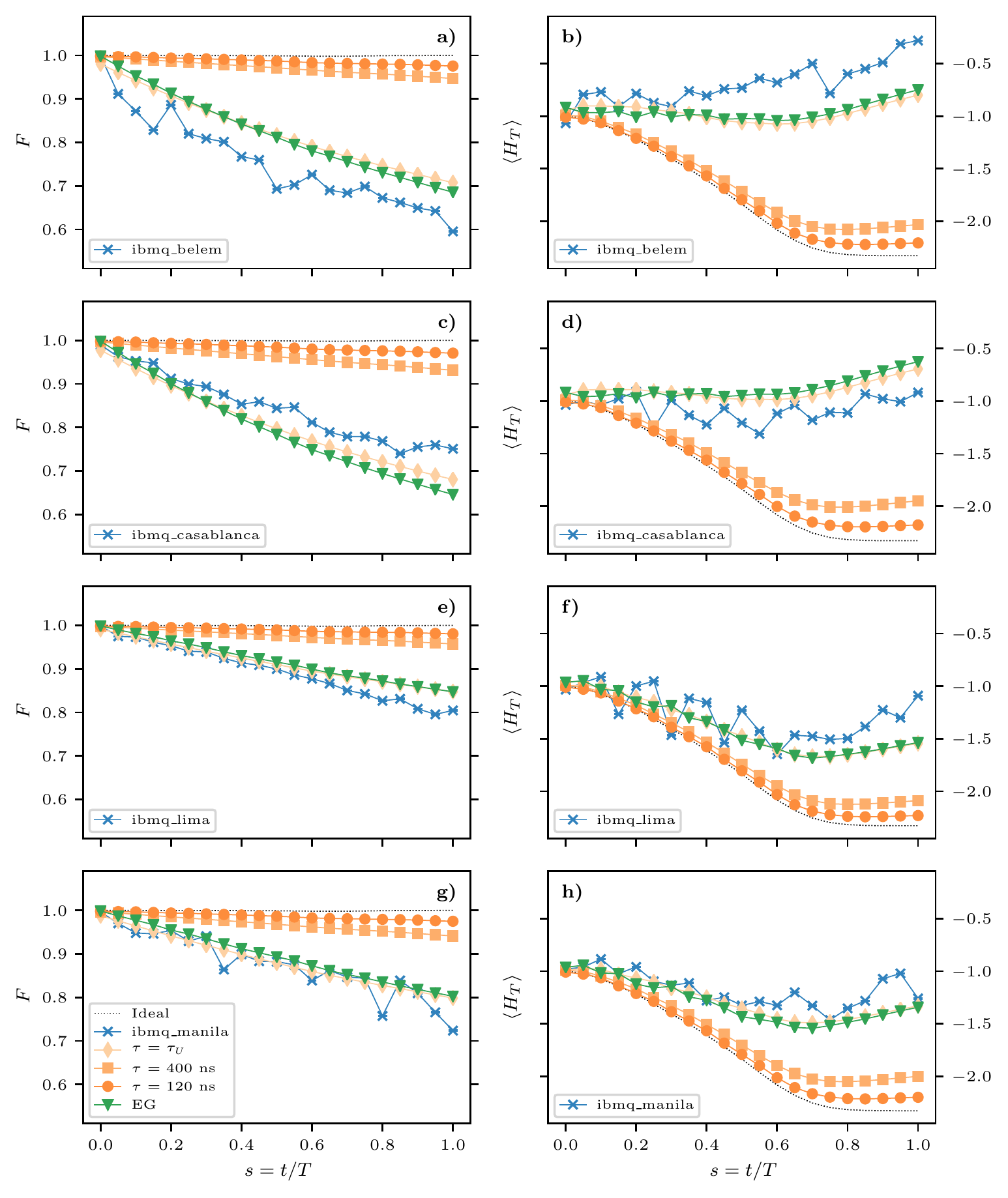}
\caption{Comparison between results from classical device-level simulations implementing the adiabatic evolution of Eq.~\eqref{eq:stp}, and those from simulation conducted on the IBMQ Belem, Casablanca, Lima and Manila systems. Panels {\bf{a)}}, {\bf{c)}}, {\bf{e)}} and {\bf{g)}} show the evolution of the instantaneous fidelity $F$. Device-level simulations implementing the adiabatic evolution through elementary or customized gates of length $\tau_U$ (green triangles and orange diamonds, respectively) reached the target state with fidelities comparable to those reached by simulations on IBMQ systems (blue crosses). Device-level simulations using shorter gates (orange squares and circles) reach fidelities around 95\%, allowing for a better extraction of the target state's spectroscopic information, as shown by the evolution of $\braket{H_T}$ in panels {\bf{b)}}, {\bf{d)}}, {\bf{f)}} and {\bf{h)}}.} 
\label{fig:results}
\end{figure*}

First, we conducted device level simulations of the adiabatic evolution of the two-spin system in which each of the short-time propagators in Eq.~\eqref{eq:stp} is realized through a single customized gate of pulse length $\tau=120$ ns. These calculations  reached the target state with fidelities around 95\%, greatly improving the results from simulations on IBMQ systems, as shown in Figs.~\ref{fig:results} {\bf{a)}}, {\bf{c)}}, {\bf{e)}}, and {\bf{g)}}. While the pulse length $\tau=120$~ns is the shortest producing gate infidelities below $10^{-4}$, the resulting pulse amplitudes surpass the arbitrary threshold $|\epsilon_{\rm I,Q}^i|<\alpha/20$ [shown as the region between dotted lines in Fig.~\ref{fig:controlpulses}], which could make the corresponding pulses unsuitable for implementation on present-day quantum devices.
Next, we performed device-level simulations implementing customized gates of length $\tau=400$~ns, comparable with the implementation times of the CNOT gate on IBMQ systems. These device-level simulations still improve over those performed on IBMQ systems, reaching the target state with fidelities around 90\%.
Finally, to enable a comparison with the IBMQ results, we performed device-level simulations with customized gates which length approximately match the implementation time of a short-time propagator on IBMQ systems, $\tau_U$. The values used for $\tau_U$ are listed on Table~\ref{table:IBMQspecs}. The fidelities obtained from these simulations range between 65\% and 85\%, in good agreement with their IBMQ counterparts.
An even more direct comparison can be made with device-level simulations in which every elementary gate in the circuit implementing the adiabatic evolution on IBMQ systems is realized with customized control pulses. To conduct these last set of simulations, we started by combining each pair of simultaneous U3 gates in the short-time propagators into two-qubit ${\rm U3}^{(2)}={\rm U3}_1{\rm U3}_2$ gates, as schematically shown in Fig.~\ref{fig:directcomparison}.
Then, we realized the CNOT and ${\rm U3}^{(2)}$ gates with control pulses the combined lengths of which approximately matches the implementation time of the short-time propagators on IBMQ systems, $\tau_U$.

In Fig.~\ref{fig:results} we compare the results from our device-level simulation to those from quantum simulations on IBMQ systems.
Results from device-level simulations implementing the adiabatic evolution through CNOT and ${\rm U3}^{(2)}$ gates, shown as green triangles, closely follow those extracted from simulations on the IBMQ systems, shown as blue crosses. The results from device-level simulations realizing the short-time propagators in Eq.~\eqref{eq:stp} in terms of single two-qubit gates with control pulses of lengths $\tau=\tau_U$, $\tau=400$~ns and $\tau=120$~ns, respectively shown as orange diamonds, squares and circles, make evident the improvement that can be achieved by implementing quantum algorithms through customized gates.

\section{Conclusions}
\label{sec:conclusions}
We presented a noise-resilient approach for quantum state preparation using a minimal set of customized quantum gates to realize the adiabatic evolution of a two-spin system. Employing a model of a two qubit processor consisting of two capacitively coupled superconducting transmons, we carried out classical device-level simulations of the adiabatic evolution implemented with customized two-qubit gates and compared them with (experimental) digital simulations on IBMQ systems. We showed that high-fidelity customized gates for the short-time propagators can be implemented with pulses of varying length, ranging from 2500 to 120 ns, where the lower limit is set by considerations related to the validity of the adopted model and the specifications of standard waveform generators.
When the implementation time of the short-time propagators are similar, the state fidelities from our device-level simulations and IBMQ are comparable, indicating that our two-qubit processor model provides a realistic description of the quantum hardware. As the length of the control pulses realizing the customized gates and, consequently, the implementation time of the adiabatic evolution decrease, the loss of coherence due to the interaction between the quantum computer and its environment is greatly reduced. The fidelities achieved through customized gates greatly improve over those resulting from implementation through elementary gates, reaching the target state with up to 95\% fidelity for control pulses with length $\tau=120$~ns.

High-fidelity quantum state preparation is essential for the subsequent accurate extraction of spectroscopic properties from the desired state. In this study, we demonstrated the extraction of the energy of the state by calculating the expectation value of the target Hamiltonian. The value extracted from our best (i.e., shortest) device-level simulation, $\braket{H_T}(T)=-2.2$, is in good agreement with the exact result of $E_T=-2.328$. This is a large improvement over the energies obtained from the IBMQ simulations (ranging in mean value from $-1.25$ to $-0.25$) and, more in general, simulations on present devices implementing adiabatic evolution through elementary gates. In the future, one can envision obtaining even an higher accuracy by employing the present approach as a preconditioner for other quantum state preparation methods, such as the recently proposed quantum imaginary-time propagation~\cite{turro_f2021} and rodeo algorithms~\cite{Choi:2020pdg}.

While in this study we restricted ourselves to the simulation of a simple two-spin system, requiring only two qubits, the use of customized gates can be scaled to more complex simulations involving a larger number of spins by approximating the short-time propagators in terms of propagators for its two-particle subsystems (two-qubit gates). Device-level simulations for such multi-particle systems will require more complex multi-qubit models with time-varying couplings to minimize unwanted crosstalk among qubits not immediately involved in the propagation. In the future, we plan to explore the adiabatic evolution of linear systems of a larger number of spins.
Given the significant advantage of using customized gates demonstrated in this study, it will be also interesting to explore adiabatic evolution on IBMQ using the recently deployed pulse-level access capability.
Finally, the significant reduction in implementation time afforded by the proposed use of customized short-time propagator gates opens also the way to noise-resilient simulations of dynamical properties, such as scattering processes, through real-time evolution following preparation with adiabatic evolution.

\begin{acknowledgments}
The authors of this paper would like to thank Jonathan L. DuBois, Yaniv J. Rosen and Alessandro Roggero for useful discussions.
This work was performed under the auspices of the U.S. Department of Energy by Lawrence Livermore National Laboratory under Contract No.\ DE-AC52-07NA27344. We gratefully acknowledge support from the Laboratory Directed Research and Development (Grant No.\ 19-DR-005). We also acknowledge financial support from the U.S. Department of Energy (DE-SC0021152 and DE-SC0013365) and the NUCLEI SciDAC-4 collaboration.  Computing support for this work came from the LLNL institutional Computing Grand Challenge program. We acknowledge the use of IBMQ services for this work. The views expressed are those of the authors, and do not reflect the official policy or position of IBM or the IBMQ team. In this paper we used \textit{ibmq\_belem}, \textit{ibmq\_casablanca}, \textit{ibmq\_lima} and \textit{ibmq\_manila}, some of the IBMQ Falcon Processors.
\end{acknowledgments}

\appendix
\section{Decomposition of a short-time propagator}
\label{sec:decomposition}
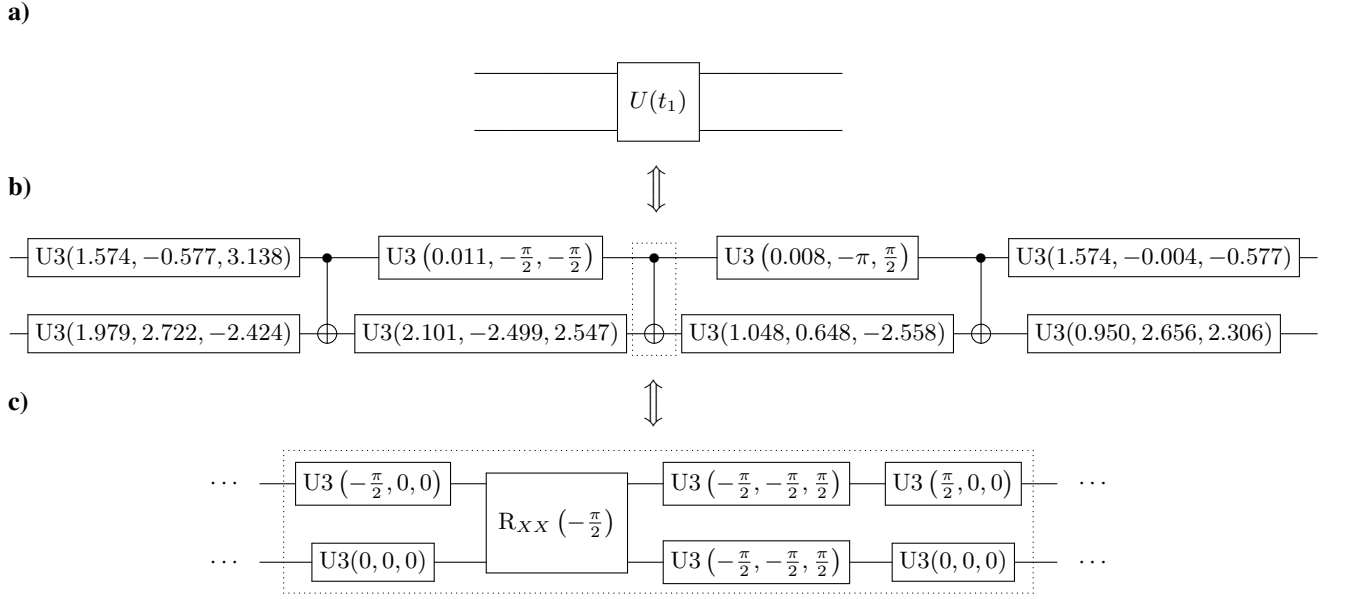
\begin{figure*}[ht]
\centering
\begin{flushleft}{\quad \normalsize{\bf{a)}}}\end{flushleft}\[
\begin{array}{c}
\Qcircuit @C=1.5em @R=1.5em {
& \qw & \qw & \qw & \multigate{1}{U(t_1)} & \qw & \qw & \qw & \qw \\
& \qw & \qw & \qw & \ghost{U(t_1)} & \qw & \qw & \qw & \qw
}
\end{array}\]
\begin{flushleft}
\quad \normalsize{\bf{b)}} \qquad\qquad\qquad\qquad\qquad
\qquad\qquad\qquad\qquad\qquad\qquad\quad $\Big\Updownarrow$
\end{flushleft}\[
\begin{array}{c}
\Qcircuit @C=0.75em @R=1.5em {
& & \gate{{\rm U3}(1.574,-0.577,3.138)} & \ctrl{1} & \gate{{\rm U3}\left(0.011,-\frac{\pi}{2},-\frac{\pi}{2}\right)} & \ctrl{1} & \gate{{\rm U3}\left(0.008,-\pi,\frac{\pi}{2}\right)} & \ctrl{1} & \gate{{\rm U3}(1.574,-0.004,-0.577)} & \qw & \\
& & \gate{{\rm U3}(1.979,2.722,-2.424)} & \targ & \gate{{\rm U3}(2.101,-2.499,2.547)} & \targ & \gate{{\rm U3}(1.048,0.648,-2.558)} & \targ & \gate{{\rm U3}(0.950,2.656,2.306)} & \qw & {}\gategroup{1}{6}{2}{6}{1.0em}{.}
}
\end{array}\]
\begin{flushleft}
\quad \normalsize{\bf{c)}}
\qquad\qquad\qquad\qquad\qquad
\qquad\qquad\qquad\qquad\qquad\qquad\quad $\Big\Updownarrow$
\end{flushleft}\[
\begin{array}{c}
\Qcircuit @C=1.5em @R=1.5em {
\cdots & & \gate{{\rm U3}\left(-\frac{\pi}{2},0,0\right)} & \multigate{1}{{\rm R}_{XX}\left(-\frac{\pi}{2}\right)} &  \gate{{\rm U3}\left(-\frac{\pi}{2},-\frac{\pi}{2},\frac{\pi}{2}\right)} & \gate{{\rm U3}\left(\frac{\pi}{2},0,0\right)} & \qw & \cdots \\
\cdots & & \gate{{\rm U3}(0,0,0)} & \ghost{{\rm R}_{XX}\left(-\frac{\pi}{2}\right)} & \gate{{\rm U3}\left(-\frac{\pi}{2},-\frac{\pi}{2},\frac{\pi}{2}\right)} & 
\gate{{\rm U3}(0,0,0)} & \qw & \cdots \gategroup{1}{3}{2}{6}{1.0em}{.}
}
\end{array}\]
\caption{Decomposition of the first
short-time propagator in
Eq.~\eqref{eq:stp}
[circuit {\bf{a)}}] into CNOT and U3 gates [circuit {\bf{b)}}]. Circuit {\bf{c)}} shows the decomposition of the CNOT gate into elementary gates.
}
\label{fig:firststpdecomp}
\end{figure*}

As mentioned in Sec.~\ref{sec:implementation}, any two-qubit gate can be decomposed into three CNOT gates where the first qubit controls the second one,
\begin{equation}
\begin{array}{c}

\Qcircuit @C=.25em @R=.7em {
& \ctrl{1} & \qw \\
& \targ  & \qw }
\end{array} = \left(\begin{array}{cccc}
1 & 0 & 0 & 0 \\
0 & 1 & 0 & 0 \\
0 & 0 & 0 & 1 \\
0 & 0 & 1 & 0
\end{array}\right),
\end{equation}
and eight U3 gates, which can be written in terms of $x$ and $z$ one-qubit rotations
\begin{equation}
\begin{aligned}
{\rm U3}(\theta,\phi,\lambda) =& R_z(\phi) R_x\left(-\frac{\pi}{2}\right) R_z(\theta) R_x\left(\frac{\pi}{2}\right) R_z(\lambda) \\
=& \left( \begin{array}{cc}
\cos\left(\frac{\theta}{2}\right) & -e^{i\lambda}\sin\left(\frac{\theta}{2}\right) \\
e^{i\phi}\sin\left(\frac{\theta}{2}\right) & e^{i(\phi+\lambda)}\cos\left(\frac{\theta}{2}\right)
\end{array} \right).
\end{aligned}
\end{equation}
In Fig.~\ref{fig:firststpdecomp} we show the quantum circuit resulting from the decomposition in terms of these gates of the first short-time propagator in Eq.~\eqref{eq:stp}, obtained using Qiskit's build-in function \textit{quantum\_info.two\_qubit\_cnot\_decompose}.

\section{Model Hamiltonian}
\label{sec:qpumodel}
In terms of the number of Cooper pairs $n$ and the flux $\phi$, the Hamiltonian of a superconducting transmon up to fourth order in $\phi$ is~\cite{koch_j2007}
\begin{equation}
\begin{aligned}
H =& 4E_C n^2 - E_J\cos(\phi)\\
\approx& 4E_C n^2 - E_J + \frac{E_J}{2}\phi^2 - \frac{E_J}{24}\phi^4,
\end{aligned}
\end{equation}
where $E_C$ and $E_J$ are respectively the energies stored in the capacitor and Josephson junction. In terms of the transmon's creation and annihilation operators, defined by
\begin{equation}
n = i\left(\frac{E_J}{32E_C}\right)^{\tfrac{1}{4}}(a^\dagger-a)
\qquad
\phi = \left(\frac{2E_C}{E_J}\right)^{\tfrac{1}{4}}(a^\dagger+a),
\end{equation}
the above Hamiltonian takes the form
\begin{equation}
\begin{aligned}
H \approx& 
\omega a^\dagger a - \frac{\alpha}{6}(a^\dagger+a)^4,
\end{aligned}
\end{equation}
where we defined $\omega\equiv(8E_CE_J)^{1/2}$ and
$\alpha\equiv E_C/2$. Using Baker-Campbell-Hausdorff formula, it can be shown that for a transformation
$U=\exp\left(-i\Omega ta^\dagger a\right)$
\begin{equation}
Ua_1\cdots a_lU^\dagger=e^{\left(-i(m-n)\Omega t\right)}a_1\cdots a_l,
\end{equation}
where  $a_1\cdots a_l$ is a chain of creation and annihilation operators (that is, $a_i\in\{a^\dagger,a\}$), and $m$ and $n$ are the number of creation and annihilation operators in the chain, respectively.
Under this transformation, after dropping constant and fast-rotating terms with $|m-n|>1$, the Hamiltonian of the transmon takes the form
\begin{equation}
\begin{aligned}
H\rightarrow H' =& UHU^\dagger + i\dot{U}U^\dagger\\
\approx& (\omega+\alpha+\Omega)a^\dagger a - \alpha a^\dagger aa^\dagger a.
\end{aligned}
\end{equation}

Now, the Hamiltonian of two capacitively coupled transmons controlled by microwave pulses can be approximately written as~\cite{jones_t2021, krantz_p2019}
\begin{equation}
\begin{aligned}
H \approx& \sum_{i=1}^2\left(4E_{C_i}n_i^2 + \frac{E_{J_i}}{2}\phi_i^2 - \frac{E_{J_i}}{24}\phi_i^4\right)
+ \frac{8E_{C_1}E_{C_2}}{E_{C_g}}n_1n_2\\
&+ 2\sum_{i=1}^2 \eta_i \left[\epsilon_{\rm I}^i(t)\sin(\Omega_it) - \epsilon_{\rm Q}^i(t)\cos(\Omega_it)\right] n_i\\
=& \sum_{i=1}^2\left[\omega_i a_i^\dagger a_i - \frac{\alpha_i}{6}(a_i^\dagger + a_i)^4\right] + g(a_1^\dagger - a_1)(a_2^\dagger - a_2)\\
&+ 2i\sum_{i=1}^2 \left[\epsilon_{\rm I}^i(t)\sin(\Omega_it) - \epsilon_{\rm Q}^i(t)\cos(\Omega_it)\right] (a_i^\dagger-a_i),
\end{aligned}
\end{equation}
where the term proportional to $g\equiv 8E_{C_1}E_{C_2}/E_{C_g}\eta_1\eta_2$ with $\eta_i\equiv(32E_{C_i}/E_{J_i})^{1/4}$ describes the interaction or crosstalk between the transmons due to its capacitive coupling.
Under the transformation
\begin{equation}
U=\exp\left(-i\Omega_1t a_1^\dagger a_1 - i\Omega_2t a_2^\dagger a_2\right),
\end{equation}
with $\Omega_i=-\omega_i-\alpha_i$, the above Hamiltonian takes the form
\begin{equation}
\begin{aligned}
H \approx& -\sum_{i=1}^2 \alpha_i a_i^\dagger a_i a_i^\dagger a_i - g\left(a_1^\dagger a_2 + a_1 a_2^\dagger \right)\\
&+ \sum_{i=1}^2 \left[\epsilon_{\rm I}^i(t)(a_i^\dagger+a_i) - i\epsilon_{\rm Q}^i(t)(a_i^\dagger-a_i)\right],
\end{aligned}
\end{equation}
after dropping constant and fast-rotating terms, and assuming that $\Omega_1\approx\Omega_2$.

\section{Quantum expectation estimation}
\label{sec:tomography}
Quantum computers are generally measured in the eigenbasis of the Pauli $Z$ matrix. Measuring an ensemble of systems in the state $\ket{\psi}$ yields the occupation probabilities $|c_{q_1q_2\ldots q_n}|^2$, where the expansion coefficients $c_{q_1q_2\ldots q_n}$ are defined by
\begin{equation}
\ket{\psi} = \sum_{q_1,q_2,\ldots,q_n\in\{0,1\}}c_{q_1q_2\ldots q_n}\ket{q_1q_2\ldots q_n}
\end{equation}

The expectation value of the two-spin operator $\sigma^z_1\sigma^z_2$ in the two-qubit state $\ket{\psi}$ can be easily extracted from experiments due to the fact that this operator is diagonal in the measurement basis
\begin{equation}
\begin{aligned}
\bra{\psi} \sigma^z_1\sigma^z_2\ket{\psi} =& \sum_{i,j,k,l\in\{0,1\}} \braket{\psi|ij} \bra{ij} \sigma^z_1\sigma^z_2 \ket{kl} \braket{kl|\psi}\\
=& \sum_{i,j\in\{0,1\}} \bra{ij} \sigma^z_1\sigma^z_2 \ket{ij} |c_{ij}|^2.
\end{aligned}
\end{equation}

Now, the expectation value of a general two-spin operator $\sigma^a_1\sigma^b_2$ with $a,b\in\{0,x,y,z\}$ ($\sigma^0_i\equiv I_i$, the identity operator) in the state $\ket{\psi}$ can also be extracted from experiments 
by means of any unitary transformation $U$ such that
\begin{equation}
U \sigma^a_1\sigma^b_2 U^\dagger = \sigma^z_1\sigma^z_2,
\end{equation}
as
\begin{equation}
\begin{aligned}
\bra{\psi} \sigma^a_1\sigma^b_2\ket{\psi} =& \bra{\psi}U^\dagger U \sigma^a_1\sigma^b_2 U^\dagger U\ket{\psi}
= \bra{\psi'} \sigma^z_1\sigma^z_2\ket{\psi'}\\
=& \sum_{i,j\in\{0,1\}} \bra{ij} \sigma^z_1\sigma^z_2 \ket{ij} |c_{ij}'|^2,
\end{aligned}
\label{eq:pauliexpectation}
\end{equation}
where
\begin{equation}
\ket{\psi'} = U\ket{\psi} = \sum_{i,j\in\{0,1\}}c_{ij}'\ket{ij}.
\end{equation}
Thus, it is possible to extract the expectation value of any Hamiltonian of the form
\begin{equation}
H = \sum_{a,b\in\{0,x,y,z\}}h_{ab}\sigma^a_1\sigma^b_2,
\end{equation}
such as the two-spin Hamiltonian of Eq.~\eqref{eq:HT}. In this work, we employed the transformations $U_x=u^x_1u^x_2$ and $U_y=u^y_1u^y_2$ with
\begin{equation}
u^x_i = {\rm U3}\left(\frac{\pi}{2},0,\pi\right)\qquad
u^y_i = {\rm U3}\left(\frac{\pi}{2},\frac{\pi}{2},\frac{\pi}{2}\right)
\end{equation}
to extract the expectation values of $\sigma^x_1\sigma^x_2$ and $\sigma^y_1\sigma^y_2$, respectively.

To approximately extract the fidelity at time $t_k$, $F(t_k)$, we take advantage of the fact that [see Eqs.~\eqref{eq:stp} and \eqref{eq:psi0}]
\begin{align}
\bra{\phi(t_k)} \approx \bra{\psi(t_k)} =& \bra{00} X^{(2)} H^{(2)} {\mathcal U}^\dagger(0,t_k)\nonumber\\
\approx& \bra{00} X^{(2)} H^{(2)} \prod_{i=k}^1 U^\dagger(t_i), 
\end{align}
decompose the overall unitary operator 
\begin{equation}
\widetilde U(t_k)=X^{(2)} H^{(2)} \prod_{i=k}^1 U^\dagger(t_i),
\end{equation}
into a single circuit with Qiskit and apply it at the end of the evolution. This yields
\begin{equation}
\begin{aligned}
F(t_k)=|\braket{\phi(t_k)|\psi(t_k)}| \approx& |\bra{00}\widetilde U(t_k)\ket{\psi(t_k)}|\\
=& |\braket{00|\widetilde\psi(t_k)}|\\
=& |\widetilde c_{00}|.
\end{aligned}
\label{eq:figexpect}
\end{equation}

\section{Measurement errors}
\label{sec:mitigation}
The measurement error mitigation scheme follows closely that employed and discussed in Ref.~\cite{roggero_a2020}. Let the probability to measure a qubit in the state $\ket{i}$ when it was prepared in state $\ket{j}$ be $p_{ij}$. The occupation probabilities measured in an experiment, arranged in the vector $\ket{c_{\rm exp}^2}$, are related to the true ones, $\ket{c_{\rm true}^2}$, through the confusion or error matrix $P$
\begin{equation}
\begin{gathered}
\ket{|c|_{\rm exp}^2} = P \ket{|c|_{\rm true}^2}\\
{\rm or}\\
\ket{|c|_{\rm true}^2} = P^{-1} \ket{|c|_{\rm exp}^2},
\end{gathered}
\end{equation}
constructed from the probabilities to measure a two-qubit system in the state $\ket{ij}$ when it was prepared in the state $\ket{kl}$, denoted by $p_{ij,kl}$.
Under the assumption that the measurement errors in qubits 1 and 2 are independent of each other, the confusion matrix can be written in terms of the single qubit measurement probabilities $p_{ij}$ as
\begin{widetext}
\begin{equation}
\begin{aligned}
P 
%
%
\approx& \left( \begin{array}{cccc}
(1-p_{10})_1 (1-p_{10})_2 &
(1-p_{10})_1 (p_{01})_2 &
(p_{01})_1 (1-p_{10})_2 &
(p_{01})_1 (p_{01})_2 \\
(1-p_{10})_1 (p_{10})_2 &
(1-p_{10})_1 (1-p_{01})_2 &
(p_{01})_1 (p_{10})_2 &
(p_{01})_1 (1-p_{01})_2 \\
(p_{10})_1 (1-p_{10})_2 &
(p_{10})_1 (p_{01})_2 &
(1-p_{01})_1 (1-p_{10})_2 &
(1-p_{01})_1 (p_{01})_2 \\
(p_{10})_1 (p_{10})_2 &
(p_{10})_1 (1-p_{01})_2 &
(1-p_{01})_1 (p_{10})_2 &
(1-p_{01})_1 (1-p_{01})_2
\end{array}\right).
\end{aligned}
\end{equation}
\end{widetext}
At the beginning of each time step, we perform two calibration experiments, preparing the device in the states $|00\rangle$ and $|11\rangle$. The measurements of these experiments correspond to the first and fourth columns of the confusion matrix. From them, we can extract the single qubit measurement probabilities, $p_{ij}$, and construct the full confusion matrix.

\section{Identification of dominant error sources}

\newcommand{\op}[1]{\ensuremath{\bm{#1}}}

\begin{figure}[hb!]
\centering
\includegraphics{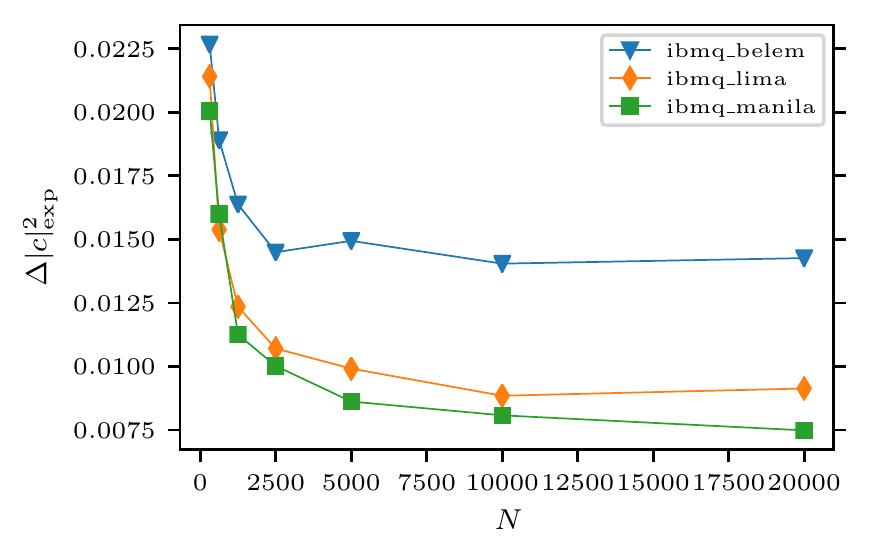}
\caption{Average error in measured occupation probabilities with the number of shots $N$. Increasing the number of shots decreases the statistical noise yielding an estimate for the quantum system's measurement error. The error in the occupation probabilities measured from simulations on IBMQ systems are bounded from above by $\delta|c|_{\rm exp}^2=0.016$, accounting for only a small fraction of the error observed in calculated fidelities and expectation values.}
\label{fig:errorwithN}
\end{figure}

There are several sources of error affecting the simulation of the adiabatic evolution performed on IBMQ hardware emulated by our classical device-level simulations. Among them we count systematic gate infidelities, stochastic dissipative processes during the circuit execution, stochastic measurement error during readout and statistical noise.
The impact of these sources on the adiabatic evolution can be studied by contrasting simulations based on Qiskit's AER Simulator, and actual simulations on IBMQ hardware. We start running simulations with Qiskit's AER simulator to measure the state
$H^{(2)}\ket{00}=\left(\ket{00}+\ket{01}+\ket{10}+\ket{11}\right)/2$, using the \textit{AerSimulator.from\_backend} method. This constructs a classical simulator with an approximate noise model for that device that includes gate errors, readout errors, and dissipative processes. Fig.~\ref{fig:errorwithN} shows the average deviation of the measured occupation probabilities from their ideal value, $1/4$, with the number of shots $N$.
As $N$ is increased, the statistical component of the error becomes negligible and the average error reaches a device-dependent plateau, that we identify with the measurement error of each IBMQ system. Results from these simulations suggest that measurement and statistical noise are not dominant sources of error, as they account for deviations from exact fidelities and expectation values (for IBMQ simulations considering $N=2500$ shots) of approximately $\delta F\approx0.01$ and $\delta\braket{H_T}\approx0.1$, much smaller from those shown in Fig.~\ref{fig:results}.

Next, we ran simulations of the adiabatic evolution with Qiskit's AER simulator choosing the \textit{``statevector''} method and saving per-shot amplitudes (this unfortunately disables the simulation of readout error). We simulate 100,000 shots for each circuit needed to compute the evolving instantaneous fidelity and the instantaneous ground state energy. Using the stored amplitudes, we compute the energy for each simulated shot following Eq.~\eqref{eq:pauliexpectation}, yielding the distribution of possible energies and repeat a similar process for the fidelity. To disentangle the impact of dissipative processes from all other sources of error, we construct the density matrix for each circuit by averaging over the shots:
\begin{align}
    \op{\rho} = \frac{1}{N_\textrm{shots}} \sum_{i\in \textrm{shots}} \ket{i}\bra{i}\,,
\end{align}
where $\ket{i}$ is the final state of the simulated circuits.
\begin{figure}[t]
\centering
\includegraphics{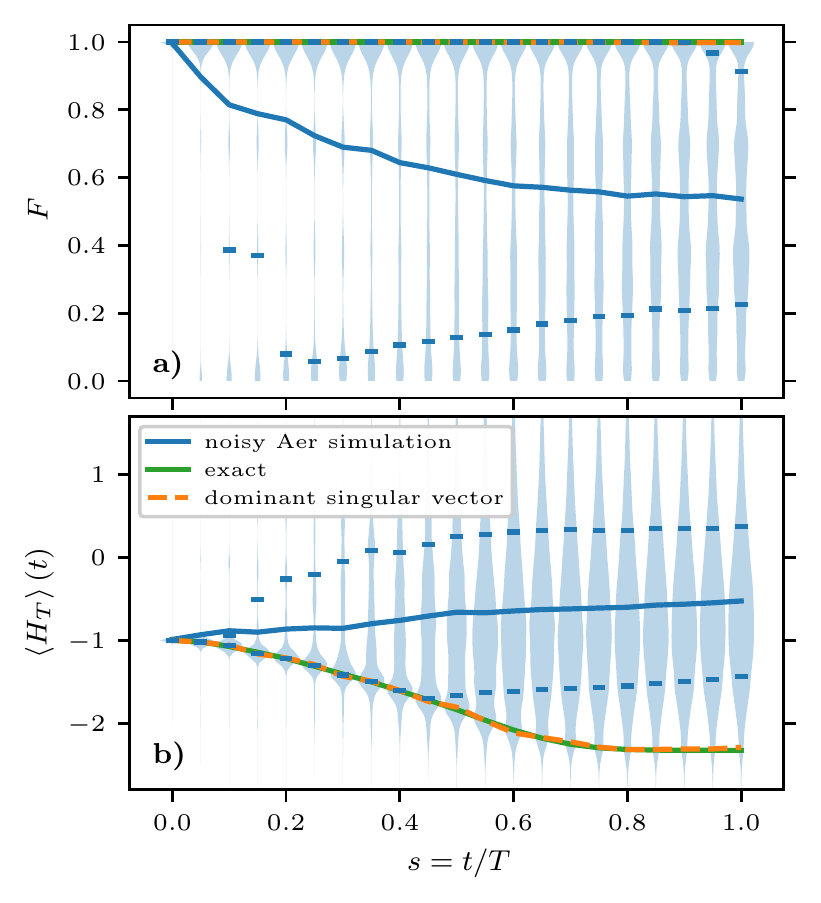}
\caption{Evolution of {\bf{a)}} the fidelity and {\bf{b)}} expectation value $\braket{H_T}$ with Qiskit's AER simulator and a noise model based on \textit{ibmq\_belem}. Solid blue lines show the expect output when including all error sources considered by AER, while the blue shaded regions indicate the distributions of likely values at each time step. Solid green lines indicate the exact answers that a perfect noiseless device would yield. Dashed orange lines result from a singular value decomposition filtering of the data underlying the blue distributions. The similarities between green and orange lines strongly suggest that dissipative processes are by far the dominant source of noise in IBMQ simulations, even very early in the evolution.}
\label{fig:noisemodel}
\end{figure}
We then compute the singular value decomposition of the density to find the most likely state, i.e., the state with largest singular value, and compute the fidelity and energy using those amplitudes. In the presence of only gate infidelity and readout errors, there should be one singular value very close to one, however the dissipative processes will produce mixed states and the largest singular value will be significantly less that its ideal value, decreasing with the circuit depth. In Fig.~\ref{fig:noisemodel} we plot the resulting distributions at each step for both {\bf{a)}} the fidelity and {\bf{b)}} the expectation value of the target Hamiltonian as blue violin plots showing the mean and $1\sigma$ quantiles. The green lines show the exact answer expected from an ideal evolution, while the orange dashed lines show results using the largest singular vector of density matrices. The difference between the dashed orange and solid blue lines results almost entirely from dissipative noise processes, which in contrast to the closeness of the dashed orange and solid green lines indicate again that other errors are extremely minor in comparison. The blue markers give a measure of the lower bound on the total uncertainty that we can anticipate from simulations on IBMQ systems.


\begin{thebibliography}{29}%
\makeatletter
\providecommand \@ifxundefined [1]{%
 \@ifx{#1\undefined}
}%
\providecommand \@ifnum [1]{%
 \ifnum #1\expandafter \@firstoftwo
 \else \expandafter \@secondoftwo
 \fi
}%
\providecommand \@ifx [1]{%
 \ifx #1\expandafter \@firstoftwo
 \else \expandafter \@secondoftwo
 \fi
}%
\providecommand \natexlab [1]{#1}%
\providecommand \enquote  [1]{``#1''}%
\providecommand \bibnamefont  [1]{#1}%
\providecommand \bibfnamefont [1]{#1}%
\providecommand \citenamefont [1]{#1}%
\providecommand \href@noop [0]{\@secondoftwo}%
\providecommand \href [0]{\begingroup \@sanitize@url \@href}%
\providecommand \@href[1]{\@@startlink{#1}\@@href}%
\providecommand \@@href[1]{\endgroup#1\@@endlink}%
\providecommand \@sanitize@url [0]{\catcode `\\12\catcode `\$12\catcode
  `\&12\catcode `\#12\catcode `\^12\catcode `\_12\catcode `\%12\relax}%
\providecommand \@@startlink[1]{}%
\providecommand \@@endlink[0]{}%
\providecommand \url  [0]{\begingroup\@sanitize@url \@url }%
\providecommand \@url [1]{\endgroup\@href {#1}{\urlprefix }}%
\providecommand \urlprefix  [0]{URL }%
\providecommand \Eprint [0]{\href }%
\providecommand \doibase [0]{http://dx.doi.org/}%
\providecommand \selectlanguage [0]{\@gobble}%
\providecommand \bibinfo  [0]{\@secondoftwo}%
\providecommand \bibfield  [0]{\@secondoftwo}%
\providecommand \translation [1]{[#1]}%
\providecommand \BibitemOpen [0]{}%
\providecommand \bibitemStop [0]{}%
\providecommand \bibitemNoStop [0]{.\EOS\space}%
\providecommand \EOS [0]{\spacefactor3000\relax}%
\providecommand \BibitemShut  [1]{\csname bibitem#1\endcsname}%
\let\auto@bib@innerbib\@empty
\bibitem [{\citenamefont {Nielsen}\ and\ \citenamefont
  {Chuang}(2010)}]{nielsen_ma2010}%
  \BibitemOpen
  \bibfield  {author} {\bibinfo {author} {\bibfnamefont {M.~A.}\ \bibnamefont
  {Nielsen}}\ and\ \bibinfo {author} {\bibfnamefont {I.~L.}\ \bibnamefont
  {Chuang}},\ }\href {\doibase 10.1017/CBO9780511976667} {\emph {\bibinfo
  {title} {{Quantum Computation and Quantum Information}}}}\ (\bibinfo
  {publisher} {Cambridge University Press},\ \bibinfo {year}
  {2010})\BibitemShut {NoStop}%
\bibitem [{\citenamefont {Steane}(1998)}]{steane_a1998}%
  \BibitemOpen
  \bibfield  {author} {\bibinfo {author} {\bibfnamefont {A.}~\bibnamefont
  {Steane}},\ }\href {\doibase 10.1088/0034-4885/61/2/002} {\bibfield
  {journal} {\bibinfo  {journal} {Rep. Prog. Phys.}\ }\textbf {\bibinfo
  {volume} {61}},\ \bibinfo {pages} {117} (\bibinfo {year} {1998})}\BibitemShut
  {NoStop}%
\bibitem [{\citenamefont {Lloyd}(1996)}]{lloyd_s1996}%
  \BibitemOpen
  \bibfield  {author} {\bibinfo {author} {\bibfnamefont {S.}~\bibnamefont
  {Lloyd}},\ }\href {\doibase 10.1126/science.273.5278.1073} {\bibfield
  {journal} {\bibinfo  {journal} {Science}\ }\textbf {\bibinfo {volume}
  {273}},\ \bibinfo {pages} {1073} (\bibinfo {year} {1996})}\BibitemShut
  {NoStop}%
\bibitem [{\citenamefont {Feynman}(1982)}]{feynman_rp1982}%
  \BibitemOpen
  \bibfield  {author} {\bibinfo {author} {\bibfnamefont {R.~P.}\ \bibnamefont
  {Feynman}},\ }\href {\doibase 10.1007/BF02650179} {\bibfield  {journal}
  {\bibinfo  {journal} {Int. J. Theor. Phys.}\ }\textbf {\bibinfo {volume}
  {21}},\ \bibinfo {pages} {467} (\bibinfo {year} {1982})}\BibitemShut
  {NoStop}%
\bibitem [{\citenamefont {Shende}\ \emph {et~al.}(2004)\citenamefont {Shende},
  \citenamefont {Markov},\ and\ \citenamefont {Bullock}}]{shende_vv2004}%
  \BibitemOpen
  \bibfield  {author} {\bibinfo {author} {\bibfnamefont {V.~V.}\ \bibnamefont
  {Shende}}, \bibinfo {author} {\bibfnamefont {I.~L.}\ \bibnamefont {Markov}},
  \ and\ \bibinfo {author} {\bibfnamefont {S.~S.}\ \bibnamefont {Bullock}},\
  }\href {\doibase 10.1103/PhysRevA.69.062321} {\bibfield  {journal} {\bibinfo
  {journal} {Phys. Rev. A}\ }\textbf {\bibinfo {volume} {69}},\ \bibinfo
  {pages} {062321} (\bibinfo {year} {2004})}\BibitemShut {NoStop}%
\bibitem [{\citenamefont {Vatan}\ and\ \citenamefont
  {Williams}(2004)}]{vatan_f2004}%
  \BibitemOpen
  \bibfield  {author} {\bibinfo {author} {\bibfnamefont {F.}~\bibnamefont
  {Vatan}}\ and\ \bibinfo {author} {\bibfnamefont {C.}~\bibnamefont
  {Williams}},\ }\href {\doibase 10.1103/PhysRevA.69.032315} {\bibfield
  {journal} {\bibinfo  {journal} {Phys. Rev. A}\ }\textbf {\bibinfo {volume}
  {69}},\ \bibinfo {pages} {032315} (\bibinfo {year} {2004})}\BibitemShut
  {NoStop}%
\bibitem [{\citenamefont {Vidal}\ and\ \citenamefont
  {Dawson}(2004)}]{vidal_g2004}%
  \BibitemOpen
  \bibfield  {author} {\bibinfo {author} {\bibfnamefont {G.}~\bibnamefont
  {Vidal}}\ and\ \bibinfo {author} {\bibfnamefont {C.~M.}\ \bibnamefont
  {Dawson}},\ }\href {\doibase 10.1103/PhysRevA.69.010301} {\bibfield
  {journal} {\bibinfo  {journal} {Phys. Rev. A}\ }\textbf {\bibinfo {volume}
  {69}},\ \bibinfo {pages} {010301} (\bibinfo {year} {2004})}\BibitemShut
  {NoStop}%
\bibitem [{\citenamefont {{Barenco, A. and Bennett, C. H. and Cleve, R. and
  DiVincenzo, D. P. and Margolus, N. and Shor, P. and Sleator, T. and Smolin,
  J. A. and Weinfurter, H.}}(1995)}]{barenco_a1995}%
  \BibitemOpen
  \bibfield  {author} {\bibinfo {author} {\bibnamefont {{Barenco, A. and
  Bennett, C. H. and Cleve, R. and DiVincenzo, D. P. and Margolus, N. and Shor,
  P. and Sleator, T. and Smolin, J. A. and Weinfurter, H.}}},\ }\href {\doibase
  10.1103/PhysRevA.52.3457} {\bibfield  {journal} {\bibinfo  {journal} {Phys.
  Rev. A}\ }\textbf {\bibinfo {volume} {52}},\ \bibinfo {pages} {3457}
  (\bibinfo {year} {1995})}\BibitemShut {NoStop}%
\bibitem [{\citenamefont {Wu}\ \emph {et~al.}()\citenamefont {Wu},
  \citenamefont {Wendt}, \citenamefont {Coello~P{\'e}rez}, \citenamefont
  {Kravvaris}, \citenamefont {Martinez}, \citenamefont {Castelli},
  \citenamefont {Rosen}, \citenamefont {Ormand}, \citenamefont {Pederiva},
  \citenamefont {DuBois},\ and\ \citenamefont {Quaglioni}}]{wu2021}%
  \BibitemOpen
  \bibfield  {author} {\bibinfo {author} {\bibfnamefont {X.}~\bibnamefont
  {Wu}}, \bibinfo {author} {\bibfnamefont {K.~A.}\ \bibnamefont {Wendt}},
  \bibinfo {author} {\bibfnamefont {E.~A.}\ \bibnamefont {Coello~P{\'e}rez}},
  \bibinfo {author} {\bibfnamefont {K.}~\bibnamefont {Kravvaris}}, \bibinfo
  {author} {\bibfnamefont {L.}~\bibnamefont {Martinez}}, \bibinfo {author}
  {\bibfnamefont {L.}~\bibnamefont {Castelli}}, \bibinfo {author}
  {\bibfnamefont {Y.~J.}\ \bibnamefont {Rosen}}, \bibinfo {author}
  {\bibfnamefont {W.~E.}\ \bibnamefont {Ormand}}, \bibinfo {author}
  {\bibfnamefont {F.}~\bibnamefont {Pederiva}}, \bibinfo {author}
  {\bibfnamefont {J.~L.}\ \bibnamefont {DuBois}}, \ and\ \bibinfo {author}
  {\bibfnamefont {S.}~\bibnamefont {Quaglioni}},\ }\href@noop {} {}\bibinfo
  {note} {In preparation}\BibitemShut {NoStop}%
\bibitem [{\citenamefont {Holland}\ \emph {et~al.}(2020)\citenamefont
  {Holland}, \citenamefont {Wendt}, \citenamefont {Kravvaris}, \citenamefont
  {Wu}, \citenamefont {Ormand}, \citenamefont {DuBois}, \citenamefont
  {Quaglioni},\ and\ \citenamefont {Pederiva}}]{holland_et2020}%
  \BibitemOpen
  \bibfield  {author} {\bibinfo {author} {\bibfnamefont {E.~T.}\ \bibnamefont
  {Holland}}, \bibinfo {author} {\bibfnamefont {K.~A.}\ \bibnamefont {Wendt}},
  \bibinfo {author} {\bibfnamefont {K.}~\bibnamefont {Kravvaris}}, \bibinfo
  {author} {\bibfnamefont {X.}~\bibnamefont {Wu}}, \bibinfo {author}
  {\bibfnamefont {W.~E.}\ \bibnamefont {Ormand}}, \bibinfo {author}
  {\bibfnamefont {J.~L.}\ \bibnamefont {DuBois}}, \bibinfo {author}
  {\bibfnamefont {S.}~\bibnamefont {Quaglioni}}, \ and\ \bibinfo {author}
  {\bibfnamefont {F.}~\bibnamefont {Pederiva}},\ }\href {\doibase
  10.1103/PhysRevA.101.062307} {\bibfield  {journal} {\bibinfo  {journal}
  {Phys. Rev. A}\ }\textbf {\bibinfo {volume} {101}},\ \bibinfo {pages}
  {062307} (\bibinfo {year} {2020})}\BibitemShut {NoStop}%
\bibitem [{\citenamefont {Born}\ and\ \citenamefont {Fock}(1928)}]{born_m1928}%
  \BibitemOpen
  \bibfield  {author} {\bibinfo {author} {\bibfnamefont {M.}~\bibnamefont
  {Born}}\ and\ \bibinfo {author} {\bibfnamefont {V.}~\bibnamefont {Fock}},\
  }\href {\doibase 10.1007/BF01343193} {\bibfield  {journal} {\bibinfo
  {journal} {{Zeitschrift f\"ur Physik}}\ }\textbf {\bibinfo {volume} {51}},\
  \bibinfo {pages} {165} (\bibinfo {year} {1928})}\BibitemShut {NoStop}%
\bibitem [{\citenamefont {Albash}\ and\ \citenamefont
  {Lidar}(2018)}]{albash_t2018}%
  \BibitemOpen
  \bibfield  {author} {\bibinfo {author} {\bibfnamefont {T.}~\bibnamefont
  {Albash}}\ and\ \bibinfo {author} {\bibfnamefont {D.~A.}\ \bibnamefont
  {Lidar}},\ }\href {\doibase 10.1103/RevModPhys.90.015002} {\bibfield
  {journal} {\bibinfo  {journal} {Rev. Mod. Phys.}\ }\textbf {\bibinfo {volume}
  {90}},\ \bibinfo {pages} {015002} (\bibinfo {year} {2018})}\BibitemShut
  {NoStop}%
\bibitem [{\citenamefont {Boixo}\ and\ \citenamefont
  {Somma}(2010)}]{boixo_s2010}%
  \BibitemOpen
  \bibfield  {author} {\bibinfo {author} {\bibfnamefont {S.}~\bibnamefont
  {Boixo}}\ and\ \bibinfo {author} {\bibfnamefont {R.~D.}\ \bibnamefont
  {Somma}},\ }\href {\doibase 10.1103/PhysRevA.81.032308} {\bibfield  {journal}
  {\bibinfo  {journal} {Phys. Rev. A}\ }\textbf {\bibinfo {volume} {81}},\
  \bibinfo {pages} {032308} (\bibinfo {year} {2010})}\BibitemShut {NoStop}%
\bibitem [{\citenamefont {Boixo}\ \emph {et~al.}(2009)\citenamefont {Boixo},
  \citenamefont {Knill},\ and\ \citenamefont {Somma}}]{boixo_s2009}%
  \BibitemOpen
  \bibfield  {author} {\bibinfo {author} {\bibfnamefont {S.}~\bibnamefont
  {Boixo}}, \bibinfo {author} {\bibfnamefont {E.}~\bibnamefont {Knill}}, \ and\
  \bibinfo {author} {\bibfnamefont {R.~D.}\ \bibnamefont {Somma}},\ }\href
  {\doibase 10.26421/QIC9.9-10-7} {\bibfield  {journal} {\bibinfo  {journal}
  {Quantum Inf. Comput.}\ }\textbf {\bibinfo {volume} {9}},\ \bibinfo {pages}
  {833} (\bibinfo {year} {2009})}\BibitemShut {NoStop}%
\bibitem [{\citenamefont {Kitaev}(1995)}]{Kitaev:1995qy}%
  \BibitemOpen
  \bibfield  {author} {\bibinfo {author} {\bibfnamefont {A.~{\relax Yu.}.}\
  \bibnamefont {Kitaev}},\ }\href@noop {} {\  (\bibinfo {year} {1995})},\
  \Eprint {http://arxiv.org/abs/quant-ph/9511026} {arXiv:quant-ph/9511026}
  \BibitemShut {NoStop}%
\bibitem [{\citenamefont {Choi}\ \emph {et~al.}(2021)\citenamefont {Choi},
  \citenamefont {Lee}, \citenamefont {Bonitati}, \citenamefont {Qian},\ and\
  \citenamefont {Watkins}}]{Choi:2020pdg}%
  \BibitemOpen
  \bibfield  {author} {\bibinfo {author} {\bibfnamefont {K.}~\bibnamefont
  {Choi}}, \bibinfo {author} {\bibfnamefont {D.}~\bibnamefont {Lee}}, \bibinfo
  {author} {\bibfnamefont {J.}~\bibnamefont {Bonitati}}, \bibinfo {author}
  {\bibfnamefont {Z.}~\bibnamefont {Qian}}, \ and\ \bibinfo {author}
  {\bibfnamefont {J.}~\bibnamefont {Watkins}},\ }\href {\doibase
  10.1103/PhysRevLett.127.040505} {\bibfield  {journal} {\bibinfo  {journal}
  {Phys. Rev. Lett.}\ }\textbf {\bibinfo {volume} {127}},\ \bibinfo {pages}
  {040505} (\bibinfo {year} {2021})},\ \Eprint
  {http://arxiv.org/abs/2009.04092} {arXiv:2009.04092 [quant-ph]} \BibitemShut
  {NoStop}%
\bibitem [{\citenamefont {Qian}\ \emph {et~al.}(2021)\citenamefont {Qian},
  \citenamefont {Watkins}, \citenamefont {Given}, \citenamefont {Bonitati},
  \citenamefont {Choi},\ and\ \citenamefont {Lee}}]{Qian:2021wya}%
  \BibitemOpen
  \bibfield  {author} {\bibinfo {author} {\bibfnamefont {Z.}~\bibnamefont
  {Qian}}, \bibinfo {author} {\bibfnamefont {J.}~\bibnamefont {Watkins}},
  \bibinfo {author} {\bibfnamefont {G.}~\bibnamefont {Given}}, \bibinfo
  {author} {\bibfnamefont {J.}~\bibnamefont {Bonitati}}, \bibinfo {author}
  {\bibfnamefont {K.}~\bibnamefont {Choi}}, \ and\ \bibinfo {author}
  {\bibfnamefont {D.}~\bibnamefont {Lee}},\ }\href@noop {} {\  (\bibinfo {year}
  {2021})},\ \Eprint {http://arxiv.org/abs/2110.07747} {arXiv:2110.07747
  [quant-ph]} \BibitemShut {NoStop}%
\bibitem [{\citenamefont {{Poulin, D. and Qarry, A. and Somma, R. and
  Verstraete, F.}}(2011)}]{poulin_d2011}%
  \BibitemOpen
  \bibfield  {author} {\bibinfo {author} {\bibnamefont {{Poulin, D. and Qarry,
  A. and Somma, R. and Verstraete, F.}}},\ }\href {\doibase
  10.1103/PhysRevLett.106.170501} {\bibfield  {journal} {\bibinfo  {journal}
  {Phys. Rev. Lett.}\ }\textbf {\bibinfo {volume} {106}},\ \bibinfo {pages}
  {170501} (\bibinfo {year} {2011})}\BibitemShut {NoStop}%
\bibitem [{\citenamefont {{IBM Quantum}}(2021)}]{ibmq_2021}%
  \BibitemOpen
  \bibfield  {author} {\bibinfo {author} {\bibnamefont {{IBM Quantum}}},\
  }\href {https://quantum-computing.ibm.com} {}\bibinfo {howpublished}
  {https://quantum-computing.ibm.com} (\bibinfo {year} {2021})\BibitemShut
  {NoStop}%
\bibitem [{\citenamefont {Anis}\ \emph {et~al.}(2021)\citenamefont {Anis} \emph
  {et~al.}}]{qiskit_2021}%
  \BibitemOpen
  \bibfield  {author} {\bibinfo {author} {\bibfnamefont {M.~S.}\ \bibnamefont
  {Anis}} \emph {et~al.},\ }\href {\doibase 10.5281/zenodo.2573505} {\enquote
  {\bibinfo {title} {{Qiskit: An open-source framework for quantum
  computing}},}\ } (\bibinfo {year} {2021})\BibitemShut {NoStop}%
\bibitem [{\citenamefont {Motzoi}\ \emph {et~al.}(2009)\citenamefont {Motzoi},
  \citenamefont {Gambetta}, \citenamefont {Rebentrost},\ and\ \citenamefont
  {Wilhelm}}]{motzoi_f2009}%
  \BibitemOpen
  \bibfield  {author} {\bibinfo {author} {\bibfnamefont {F.}~\bibnamefont
  {Motzoi}}, \bibinfo {author} {\bibfnamefont {J.~M.}\ \bibnamefont
  {Gambetta}}, \bibinfo {author} {\bibfnamefont {P.}~\bibnamefont
  {Rebentrost}}, \ and\ \bibinfo {author} {\bibfnamefont {F.~K.}\ \bibnamefont
  {Wilhelm}},\ }\href {\doibase 10.1103/PhysRevLett.103.110501} {\bibfield
  {journal} {\bibinfo  {journal} {Phys. Rev. Lett.}\ }\textbf {\bibinfo
  {volume} {103}},\ \bibinfo {pages} {110501} (\bibinfo {year}
  {2009})}\BibitemShut {NoStop}%
\bibitem [{\citenamefont {{N. Khaneja and T. Reiss and C. Kehlet and T.
  Schulte-Herbr\"uggen and S. J. Glaser}}(2005)}]{khaneja_n2005}%
  \BibitemOpen
  \bibfield  {author} {\bibinfo {author} {\bibnamefont {{N. Khaneja and T.
  Reiss and C. Kehlet and T. Schulte-Herbr\"uggen and S. J. Glaser}}},\ }\href
  {\doibase 10.1016/j.jmr.2004.11.004} {\bibfield  {journal} {\bibinfo
  {journal} {J. Magn. Reson.}\ }\textbf {\bibinfo {volume} {172}},\ \bibinfo
  {pages} {296} (\bibinfo {year} {2005})}\BibitemShut {NoStop}%
\bibitem [{\citenamefont {Johansson}\ \emph {et~al.}(2013)\citenamefont
  {Johansson}, \citenamefont {Nation},\ and\ \citenamefont
  {Nori}}]{johansson_jr2013}%
  \BibitemOpen
  \bibfield  {author} {\bibinfo {author} {\bibfnamefont {J.~R.}\ \bibnamefont
  {Johansson}}, \bibinfo {author} {\bibfnamefont {P.~D.}\ \bibnamefont
  {Nation}}, \ and\ \bibinfo {author} {\bibfnamefont {F.}~\bibnamefont
  {Nori}},\ }\href {\doibase doi.org/10.1016/j.cpc.2012.11.019} {\bibfield
  {journal} {\bibinfo  {journal} {Comput. Phys. Commun.}\ }\textbf {\bibinfo
  {volume} {184}},\ \bibinfo {pages} {1234} (\bibinfo {year}
  {2013})}\BibitemShut {NoStop}%
\bibitem [{\citenamefont {Johansson}\ \emph {et~al.}(2012)\citenamefont
  {Johansson}, \citenamefont {Nation},\ and\ \citenamefont
  {Nori}}]{johansson_jr2012}%
  \BibitemOpen
  \bibfield  {author} {\bibinfo {author} {\bibfnamefont {J.~R.}\ \bibnamefont
  {Johansson}}, \bibinfo {author} {\bibfnamefont {P.~D.}\ \bibnamefont
  {Nation}}, \ and\ \bibinfo {author} {\bibfnamefont {F.}~\bibnamefont
  {Nori}},\ }\href {\doibase doi.org/10.1016/j.cpc.2012.02.021} {\bibfield
  {journal} {\bibinfo  {journal} {Comput. Phys. Commun.}\ }\textbf {\bibinfo
  {volume} {183}},\ \bibinfo {pages} {1760} (\bibinfo {year}
  {2012})}\BibitemShut {NoStop}%
\bibitem [{\citenamefont {{Turro}}\ \emph {et~al.}(2021)\citenamefont
  {{Turro}}, \citenamefont {{Amitrano}}, \citenamefont {{Luchi}}, \citenamefont
  {{Roggero}}, \citenamefont {{Wendt}}, \citenamefont {{DuBois}}, \citenamefont
  {{Quaglioni}},\ and\ \citenamefont {{Pederiva}}}]{turro_f2021}%
  \BibitemOpen
  \bibfield  {author} {\bibinfo {author} {\bibfnamefont {F.}~\bibnamefont
  {{Turro}}}, \bibinfo {author} {\bibfnamefont {V.}~\bibnamefont {{Amitrano}}},
  \bibinfo {author} {\bibfnamefont {P.}~\bibnamefont {{Luchi}}}, \bibinfo
  {author} {\bibfnamefont {A.}~\bibnamefont {{Roggero}}}, \bibinfo {author}
  {\bibfnamefont {K.~A.}\ \bibnamefont {{Wendt}}}, \bibinfo {author}
  {\bibfnamefont {J.~L.}\ \bibnamefont {{DuBois}}}, \bibinfo {author}
  {\bibfnamefont {S.}~\bibnamefont {{Quaglioni}}}, \ and\ \bibinfo {author}
  {\bibfnamefont {F.}~\bibnamefont {{Pederiva}}},\ }\href@noop {} {\  (\bibinfo
  {year} {2021})},\ \Eprint {http://arxiv.org/abs/2102.12260} {arXiv:2102.12260
  [quant-ph]} \BibitemShut {NoStop}%
\bibitem [{\citenamefont {Koch}\ \emph {et~al.}(2007)\citenamefont {Koch},
  \citenamefont {Yu}, \citenamefont {Gambetta}, \citenamefont {Houck},
  \citenamefont {Schuster}, \citenamefont {Majer}, \citenamefont {Blais},
  \citenamefont {Devoret}, \citenamefont {Girvin},\ and\ \citenamefont
  {Schoelkopf}}]{koch_j2007}%
  \BibitemOpen
  \bibfield  {author} {\bibinfo {author} {\bibfnamefont {J.}~\bibnamefont
  {Koch}}, \bibinfo {author} {\bibfnamefont {T.~M.}\ \bibnamefont {Yu}},
  \bibinfo {author} {\bibfnamefont {J.}~\bibnamefont {Gambetta}}, \bibinfo
  {author} {\bibfnamefont {A.~A.}\ \bibnamefont {Houck}}, \bibinfo {author}
  {\bibfnamefont {D.~I.}\ \bibnamefont {Schuster}}, \bibinfo {author}
  {\bibfnamefont {J.}~\bibnamefont {Majer}}, \bibinfo {author} {\bibfnamefont
  {A.}~\bibnamefont {Blais}}, \bibinfo {author} {\bibfnamefont {M.~H.}\
  \bibnamefont {Devoret}}, \bibinfo {author} {\bibfnamefont {S.~M.}\
  \bibnamefont {Girvin}}, \ and\ \bibinfo {author} {\bibfnamefont {R.~J.}\
  \bibnamefont {Schoelkopf}},\ }\href {\doibase 10.1103/PhysRevA.76.042319}
  {\bibfield  {journal} {\bibinfo  {journal} {Phys. Rev. A}\ }\textbf {\bibinfo
  {volume} {76}},\ \bibinfo {pages} {042319} (\bibinfo {year}
  {2007})}\BibitemShut {NoStop}%
\bibitem [{\citenamefont {{Jones}}\ \emph {et~al.}(2021)\citenamefont
  {{Jones}}, \citenamefont {{Steven}}, \citenamefont {{Poncini}}, \citenamefont
  {{Rose}},\ and\ \citenamefont {{Fedorov}}}]{jones_t2021}%
  \BibitemOpen
  \bibfield  {author} {\bibinfo {author} {\bibfnamefont {T.}~\bibnamefont
  {{Jones}}}, \bibinfo {author} {\bibfnamefont {K.}~\bibnamefont {{Steven}}},
  \bibinfo {author} {\bibfnamefont {X.}~\bibnamefont {{Poncini}}}, \bibinfo
  {author} {\bibfnamefont {M.}~\bibnamefont {{Rose}}}, \ and\ \bibinfo {author}
  {\bibfnamefont {A.}~\bibnamefont {{Fedorov}}},\ }\href
  {https://arxiv.org/abs/2102.09721} {\bibfield  {journal} {\bibinfo  {journal}
  {arXiv e-prints}\ ,\ \bibinfo {eid} {arXiv:2102.09721}} (\bibinfo {year}
  {2021})}\BibitemShut {NoStop}%
\bibitem [{\citenamefont {Krantz}\ \emph {et~al.}(2019)\citenamefont {Krantz},
  \citenamefont {Kjaergaard}, \citenamefont {Yan}, \citenamefont {Orlando},
  \citenamefont {Gustavsson},\ and\ \citenamefont {Oliver}}]{krantz_p2019}%
  \BibitemOpen
  \bibfield  {author} {\bibinfo {author} {\bibfnamefont {P.}~\bibnamefont
  {Krantz}}, \bibinfo {author} {\bibfnamefont {M.}~\bibnamefont {Kjaergaard}},
  \bibinfo {author} {\bibfnamefont {F.}~\bibnamefont {Yan}}, \bibinfo {author}
  {\bibfnamefont {T.~P.}\ \bibnamefont {Orlando}}, \bibinfo {author}
  {\bibfnamefont {S.}~\bibnamefont {Gustavsson}}, \ and\ \bibinfo {author}
  {\bibfnamefont {W.~D.}\ \bibnamefont {Oliver}},\ }\href {\doibase
  10.1063/1.5089550} {\bibfield  {journal} {\bibinfo  {journal} {Appl. Phys.
  Rev.}\ }\textbf {\bibinfo {volume} {6}},\ \bibinfo {pages} {021318} (\bibinfo
  {year} {2019})}\BibitemShut {NoStop}%
\bibitem [{\citenamefont {Roggero}\ \emph {et~al.}(2020)\citenamefont
  {Roggero}, \citenamefont {Gu}, \citenamefont {Baroni},\ and\ \citenamefont
  {Papenbrock}}]{roggero_a2020}%
  \BibitemOpen
  \bibfield  {author} {\bibinfo {author} {\bibfnamefont {A.}~\bibnamefont
  {Roggero}}, \bibinfo {author} {\bibfnamefont {C.}~\bibnamefont {Gu}},
  \bibinfo {author} {\bibfnamefont {A.}~\bibnamefont {Baroni}}, \ and\ \bibinfo
  {author} {\bibfnamefont {T.}~\bibnamefont {Papenbrock}},\ }\href {\doibase
  10.1103/PhysRevC.102.064624} {\bibfield  {journal} {\bibinfo  {journal}
  {Phys. Rev. C}\ }\textbf {\bibinfo {volume} {102}},\ \bibinfo {pages}
  {064624} (\bibinfo {year} {2020})}\BibitemShut {NoStop}%
\end{thebibliography}

%

\end{document}